\documentclass[11pt,a4paper]{article}
\usepackage{jheppub}

\usepackage{slashed}
\usepackage{verbatim}
\usepackage{latexsym}
\usepackage{euscript}
\usepackage{amsfonts}
\usepackage{verbatim}
\usepackage[]{array}
\usepackage[]{mathrsfs}
\usepackage{graphicx}
\usepackage{amsmath}
\usepackage{amssymb}
\usepackage{amsxtra}

\def\be{\begin{eqnarray}}
\def\ee{\end{eqnarray}}
\def\0{\nonumber}

\def\Tr{{\rm Tr}} 
\def\tr{{\rm tr}}

\normalfont\large

\newcommand{\ac}[2]{\{ #1 \stackrel{\star}{,} #2\}}
\newcommand{\co}[2]{\left[ #1 \stackrel{\star}{,} #2\right]}

\preprint{{ZTF-EP-21-03}}

\title{Gauging the higher-spin-like symmetries by the Moyal product}

\author[a]{M.~Cvitan,} 
\author[b]{P.~Dominis Prester,} 
\author[c]{S.~Giaccari,} 
\author[b]{M.~Pauli\v{s}i\'{c}} 
\author[b]{and I.~Vukovi\'{c}}

\affiliation[a]{Department of Physics, Faculty of Science, University
of Zagreb,\\ 
Bijeni\v{c}ka cesta 32, 10000 Zagreb, Croatia}
\affiliation[b]{Department of Physics, University of Rijeka,\\
Radmile Matej\v{c}i\'{c} 2, 51000 Rijeka, Croatia}
\affiliation[c]{Department of Sciences,
Holon Institute of Technology (HIT),\\
52 Golomb St., Holon 5810201, Israel}

\emailAdd{mcvitan@phy.hr} 
\emailAdd{pprester@phy.uniri.hr}
\emailAdd{stefanog@hit.ac.il}
\emailAdd{mateo.paulisic@phy.uniri.hr} 
\emailAdd{ivan.vukovic@phy.uniri.hr}

\abstract{
We analyze a novel approach to gauging rigid higher derivative (higher spin) symmetries of free relativistic actions defined on flat spacetime, building on the formalism originally developed by Bonora et al. and Bekaert et al. in their studies of linear coupling of matter fields to an infinite tower of higher spin fields. The off-shell definition is based on fields defined on a $2d$-dimensional master space equipped with a symplectic structure, where the infinite dimensional Lie algebra of gauge transformations is given by the Moyal commutator. Using this algebra we construct well-defined weakly non-local actions, both in the gauge and the matter sector, by mimicking the Yang-Mills procedure. The theory allows for a description in terms of an infinite tower of higher spin spacetime fields only on-shell. Interestingly, Euclidean theory allows for such a description also off-shell. 
Owing to its formal similarity to non-commutative field theories, the formalism allows for the introduction of a covariant potential which plays the role of the generalised vielbein. This covariant formulation uncovers the existence of other phases and shows that the theory can be written in a matrix model form. The symmetries of the theory are analyzed and conserved currents are explicitly constructed. By studying the spin-2 sector we show that the emergent geometry is closely related to teleparallel geometry, in the sense that the induced linear connection is opposite to Weitzenb\"{o}ck's.
}

\arxivnumber{2102.09254}

\begin{document}

\maketitle

\section{Introduction}
\label{sec:intro}

In the recent period we have witnessed a revival of interest in studying properties of gauge quantum field theories. Taking into account that it is a mathematical framework underlying our current fundamental description of Nature (via the Standard Model of elementary particles), the interest is hardly surprising. A contributive factor to this shift of interests is that frameworks invented to solve long standing issues present in the high energy regime, in particular to quantize gravity (such as string theory), are after many years of development still far from fulfilling their promise. There is a growing feeling that fresh new ideas are needed and these may arise from efforts to improve our understanding of quantum field theories by relaxing some of the standard postulates.

The well-known but still poorly understood property of all free relativistic field theories on Minkowski spacetime, is that they possess an infinite tower of abelian rigid symmetries with a corresponding tower of conserved currents represented by Lorentz tensor fields of all ranks $s\ge1$ (to which one usually refers to as spin-$s$ currents).\footnote{Likewise, in supersymmetric free field theories there is also an infinite tower of half-integer spin symmetries and corresponding currents. We shall not discuss half-integer spin symmetries in this paper.} As the rigid symmetry variations and corresponding conserved currents contain a number of derivatives growing with $s$, one can speak of higher derivative or higher spin (HS) symmetries.\footnote{We shall use both terms interchangeably, depending on the context.} The low spin sector ($s\le2$) of these symmetries plays an immensely important role, as spin-1 describes internal symmetries and gives conservation of electric or non-Abelian charges while spin-2 describes translations in spacetime and leads to conservation of energy and momentum. Completing the theories by making the rigid low spin symmetries local one obtains two pillars of our current description of the fundamental forces in Nature, Maxwell and Yang-Mills theory in the spin-1 case and theories intimately related to General Relativity in the spin-2 case.\footnote{See \cite{Hehl:2019csx} for recent review and developments.} While it is easy to dismiss HS symmetries ($s>2$) due to their higher-derivative nature, it is somewhat surprising that they do not have any role in the current description. This may be an indication that we are missing something related to higher derivative symmetries in our understanding of the QFT formalism and maybe even in our fundamental model of Nature. It is not inconceivable that  the degrees of freedom associated with these symmetries may account for dark matter, a missing puzzle in the standard cosmological model.

The idea of gauging higher derivative symmetries is not new, and is at the core of the higher spin theory program (e.g., see reviews \cite{Sorokin:2004ie,Bekaert:2005vh,Bengtsson2008,Didenko:2014dwa,Rahman:2015pzl} and references therein). The standard approach is to introduce higher spin massless fields and construct higher spin gauge QFTs. While it is easy to write equations for free higher spin fields, construction of consistent interacting QFTs containing fundamental fields of spin higher than two in Minkowski spacetime is an unsolved problem of theoretical physics.\footnote{Though there were some interesting developments recently, see e.g. \cite{Skvortsov:2020gpn,Skvortsov:2020wtf,Ponomarev:2017nrr,Ponomarev:2016lrm}.} There are several no-go theorems and obstructions which put strong restrictions on such constructions.\footnote{See \cite{BBvD,Metsaev:1991mt,Ponomarev:2016lrm,Taronna:2017wbx,Roiban:2017iqg} and reviews \cite{Bekaert-nogo,Rahman:2015pzl}.} In $d\ge4$ spacetime dimensions one needs to have an infinite tower of higher spin fields. This by itself may be beneficial as degrees of freedom in such a tower can conspire to produce softer UV behaviour of amplitudes, which is an attractive possibility. The fact that massless HS fields are not observed could be explained by spontaneous symmetry breaking such that it makes all higher spin particles massive. Another constraint is that such theories apparently must possess strongly non-local interactions \cite{Taronna:2017wbx}.\footnote{One can avoid traditional no-go theorems by taking AdS spacetime as a background, allowing for Vasiliev higher spin program \cite{Vasiliev,Bekaert:2005vh} to be developed, but so far only with partial success as no off-shell formulation of a consistent interacting HS QFT in $d\ge4$ exists. To this we add that non-locality appears to be an essential feature also in the AdS case \cite{Sleight:2017pcz}.} To this we add that in flat spacetime even the free HS theory has strongly non-local or higher-derivative field equations when the HS symmetry is unconstrained \cite{Bekaert:2003az,Francia:2007qt,Francia:2007ee,Sagnotti:2010jt,Campoleoni:2012th}, a property further studied by induced action methods in \cite{Bonora:2016otz,Bonora:2017ykb}. All in all, it appears that gauging higher derivative symmetries in general requires abandoning the condition of the strong locality of the action. 

In \cite{paper1,paper2} a different formalism for gauging higher derivative symmetries in Minkowski spacetime was proposed. It was based on the observation from \cite{Bekaert:2009ud,Bekaert:2010ky} that if one introduces interactions in otherwise free Klein-Gordon theory by coupling an infinite tower of external HS fields linearly to the conserved currents of the free theory, the resulting theory has local higher derivative symmetries whose algebra can be elegantly written by introducing an auxiliary space and using Moyal product acting on master fields defined on spacetime and the auxiliary space. The key ingredient was provided in \cite{BCDGPS} where it was shown that a similar construction can be done starting from the free Dirac theory, resulting in the same algebra of local higher derivative symmetries. In this case one directly obtains the master field gauge potential, allowing for the construction of a Yang-Mills-like action for the gauge potential master field by mimicking standard Yang-Mills prescription, including the BRST quantization rules (at least on the formal level) \cite{paper2}. The outstanding feature of the construction is that one obtains a theory defined off-shell, with the local symmetry obtained by gauging rigid higher spin symmetries of matter, in Minkowski spacetime. We shall refer to this framework as Moyal-higher-spin (MHS). In \cite{paper2} it was shown that an attempt to write the MHS Yang-Mills (MHSYM) theory as an off-shell theory of a tower of spacetime HS fields, by making reductions of the configuration space and scaling the coupling constant with an infinite volume factor, faces challenges of projecting out ghosts and preserving unitarity. We stress here that reductions considered in \cite{paper2} break the original form of the MHS symmetry, without assurance that enough of symmetry survives to guarantee unitarity, which means that the troubles should not be ascribed to the MHS theory in its original form, but to the procedure which basically changes the original theory. 
 
Instead of trying to achieve the standard HS description in terms of finite spin spacetime fields, in this paper we propose to take the MHS theory as it is naturally defined, which is a theory fundamentally described by master fields, and explore its properties. In section \ref{sec:hsym} we review the MHS construction and state some basic properties of the MHSYM theory. In particular, we show that MHSYM theory has a perturbatively stable vacuum and does not possess Ostrogradsky ghosts, and make connection with standard non-commutative YM theory. We explain why the on-shell description in terms of an infinite tower of finite spin spacetime fields does not have a non-singular off-shell formulation which keeps manifest symmetry in the Lorentzian case, but does have such a description in the Euclidean case. In section \ref{sec:hsgeom} we show that the introduction of the MHS covariant potential allows for a more general description of the MHS formalism which bears resemblance to teleparallel geometry. In section \ref{sec:mbflt} we study in more detail MHSYM theory and its generalisations, analyse its symmetries, conserved currents and phases. In section \ref{sec:mattHS} we analyse coupling to the matter sector and show that beside standard spacetime matter fields one can introduce matter in the form of master fields.  
In section \ref{sec:lows} we study the low spin sector of the expansion in HS fields and show that the spin-2 component has a differential geometric interpretation with the linear connection being opposite to Weitzenb\"{o}ck's connection. Finally, a few technical appendices are added at the end.

\section{Moyal-higher-spin formalism in the Yang-Mills formulation}
\label{sec:hsym}

\subsection{From rigid higher spin transformations to local symmetry}
\label{ssec:frigncms}

It is known that actions of free field theories in flat $d$-dimensional Minkowski spacetime $\mathcal{M}$ are invariant under infinitesimal rigid transformations of the form
\be \label{rhss}
\delta_\varepsilon \phi(x) = \sum_{n=0}^\infty (-i)^{n+1} \varepsilon^{\mu_1\ldots\mu_n}\,
 \partial_{\mu_1} \cdots \partial_{\mu_n} \phi(x) 
\ee
where symmetric Lorentz tensors $\varepsilon^{\mu_1\ldots\mu_n}$ are infinitesimal rigid (constant in spacetime) real parameters. This is an infinite dimensional Abelian symmetry, which gives the rigid $U(1)$ symmetry for $n=0$, and spacetime translations for $n=1$. For $n>1$ one speaks of (infinitesimal) rigid higher-spin (HS) transformations corresponding to spin $s=n+1$. In the case when $\phi$ is real, one has only terms with odd $n$ (even spin). With a slight abuse of language we shall refer to the whole tower as HS transformations and HS symmetries, and refer to its $s\le2$ sector as the low-spin transformations/symmetries.

Gauging of low-spin symmetries is at the core of our current description of fundamental forces and it is thus natural to explore possibilities for gauging the full tower of HS symmetries. As previous attempts based on constructing consistent interacting local HS field theories in Minkowski background basically failed and at best gave us no-go theorems, we propose a different route here. It is built upon the idea originally introduced in the case of a Klein-Gordon field in \cite{Bekaert:2009ud,Bekaert:2010ky}, which was extended for the case of a Dirac field in \cite{BCDGPS} and further developed in \cite{paper2}. In a way, our construction provides an answer to a question stated in \cite{Bekaert:2007mi}, whether a Yang-Mills-like method of building a gauge sector dynamics is possible in such an approach. By introducing a $d$-dimensional auxiliary space $\mathcal{U}$, with coordinates denoted by $u_\mu$, and borrowing from the Wigner-Weyl phase space formalism, we write free field actions in the master space as $\mathcal{M} \times \mathcal{U}$ 
\be
S_0[\phi] = \int d^dx\, d^du\, W_\phi(x,u) \star K(u) 
\ee
where $K(u)$ is the kinetic function and $\star$ is the Moyal product defined by\footnote{The most important properties of the Moyal product relevant for us are summarized in the appendix \ref{app:moyal}.}
\be \label{Moyalpr}
a(x,u) \star b(x,u) = a(x,u)\,
\exp\!\left[\frac{i}{2}\left(\stackrel{\leftarrow}{\partial}_x \!\cdot\! \stackrel{\rightarrow}{\partial}_u
 - \stackrel{\rightarrow}{\partial}_x \!\cdot\! \stackrel{\leftarrow}{\partial}_u
\right)\right] b(x,u) \;.
\ee
The rigid HS variation (\ref{rhss}) of the Wigner function then takes the following form
\be \label{Wfmhsr}
\delta_\varepsilon W_\phi(x,u) = - i [ \varepsilon(u) \stackrel{\star}{,} W_\phi(x,u) ]
\ee
where
\be \label{gHsse}
\varepsilon(u) = \sum_{n=0}^\infty \varepsilon^{\mu_1\ldots\mu_n}\,
 u_{\mu_1} \ldots u_{\mu_n} \;.
\ee
The details of the construction can be found in appendix \ref{app:Bekaert} and section \ref{ssec:framelth}. 

The advantage of the master space description comes from the properties of the Moyal product: associativity, Leibniz identity, Jacobi identity, hermicity, and cyclicity of trace. This allows us to gauge the rigid HS symmetry in the spirit of the standard Yang-Mills (YM) procedure, by promoting the HS parameter to a master field
\be
\varepsilon(u) \to \varepsilon(x,u)
\ee
while keeping the Moyal product structure by which local HS symmetry is represented. To avoid confusion with the existing literature we shall refer to this construction as the MHS symmetry. There are two obvious representations, adjoint and fundamental. In the adjoint representation MHS variations act as
\be
\delta_\varepsilon A(x,u) = - i\, [ \varepsilon(x,u) \stackrel{\star}{,} A(x,u) ]
\ee
where $A(x,u)$ can be either a composite master field, as in the case of matter described by standard spacetime fields, in which case $A(x,u)$ is the Wigner function\footnote{In this case matter field $\phi(x)$ transforms in the Hilbert space representation, see section \ref{ssec:framelth}.} $W_\phi(x,u)$, or an elementary master field. We shall refer to an object transforming in the adjoint representation as an MHS tensor. In the fundamental representation MHS variations act as
\be \label{rhsm}
\delta_\varepsilon \chi(x,u) = - i\, \varepsilon(x,u) \star \chi(x,u)
\ee
which can be applied only to complex master fields.

Finite (large) MHS transformations in the fundamental representation are
\be \label{hslloc}
\phi^\mathcal{E}(x,u) = e_\star^{-i\, \mathcal{E}(x,u)} \star \chi(x,u)
\ee
and in the adjoint representation are
\be \label{hslloca}
A^g(x,u) \equiv A^\mathcal{E}(x,u) = e_\star^{-i\, \mathcal{E}(x,u)} \star A(x,u) \star e_\star^{i\, \mathcal{E}(x,u)} \;.
\ee
The group multiplication is represented by the Moyal product. The Baker-Campbell-Hausdorff lemma guarantees that a real function $\overline{\mathcal{E}}(x,u)$ exists such that
\be
e_\star^{-i \overline{\mathcal{E}}(x,u)} = e_\star^{-i\, \mathcal{E}_1(x,u)} \star e_\star^{-i\, \mathcal{E}_2(x,u)}
\ee
for any real $\mathcal{E}_1(x,u)$ and $\mathcal{E}_2(x,u)$, which means that the group of local MHS transformations is closed. The inverse is given by
\be \label{hsinv}
\left( e_\star^{-i\, \mathcal{E}(x,u)} \right)^{-1} = e_\star^{i\, \mathcal{E}(x,u)} \;.
\ee

By making transformations local we have changed the nature of the symmetry group, from abelian to non-abelian. The Lie algebra structure of local MHS transformations is given by the Moyal bracket
\be
[ \delta_{\varepsilon_2} , \delta_{\varepsilon_1} ] = \delta_{i [\varepsilon_1 \stackrel{\star}{,} \varepsilon_2]} \;.
\ee
Taking into account that the Moyal bracket satisfies the Jacobi identity, it follows that the MHS transformations possess an infinite dimensional non-abelian Lie algebra structure.

We shall return to the detailed study of the matter sector in section \ref{sec:mattHS}. We now turn our focus to the gauge sector.

\subsection{The MHS potential and field strength}
\label{ssec:mhsstr}

Requiring a gauge symmetry necessitates the introduction of a respective connection. The algebraic properties of the Moyal product allow us to repeat the standard steps of the Yang-Mills (YM) construction. The first step is to introduce an MHS (master) gauge potential $\mathbf{h}(x)$, which is a Lie algebra valued 1-form on spacetime $\mathcal{M}$. The potential represents the connection, which generally transforms under gauge transformations as
\be
\mathbf{h}^g(x) = g(x) \mathbf{h}(x) g(x)^{-1} - i g(x) \mathbf{d} g(x)^{-1}
\ee
where $\mathbf{d}$ is the exterior derivative. In the case of MHS transformations this implies
\be \label{haHSt}
h^\mathcal{E}_a(x,u) = e_\star^{-i\, \mathcal{E}(x,u)} \star h_a(x,u) \star e_\star^{i\, \mathcal{E}(x,u)}
 - i e_\star^{-i\, \mathcal{E}(x,u)} \star \partial^x_a e_\star^{i\, \mathcal{E}(x,u)} \;.
\ee
In the YM-like description on the Minkowski spacetime there is no difference between Latin indices $a,b,\ldots$ and the Greek indices $\mu,\nu,\ldots$, so, e.g., $\partial^x_a = \delta_a^\mu \partial^x_\mu$, $h_a = \delta_a^\mu h_\mu$ and $u_a = \delta_a^\mu u_\mu$. However, we shall later show that the Taylor expansion of master fields around $u=0$ leads to an induced differential geometry interpretation in which Latin indices behave as frame indices.

For (infinitesimal) MHS variations, parameterized by $\mathcal{E}(x,u) = \varepsilon(x,u)$, 
(\ref{haHSt}) becomes
\be \label{hsvha}
\delta_\varepsilon h_a(x,u) = \partial_a^x \varepsilon(x,u) + i\, [h_a(x,u) \stackrel{\star}{,} \varepsilon(x,u)] 
\ee
which is the variation obtained in \cite{BCDGPS} by studying a Dirac field linearly coupled to an infinite tower of HS fields.

Having the potential we can introduce the MHS covariant exterior derivative
\be \label{hsced}
\mathbf{D}^\star = \mathbf{d} + i\, \mathbf{h}\: \hat{\star}
\ee 
where $\hat{\star}$ stands for the Moyal-wedge product. We can write (\ref{hsvha}) as
\be
\delta_\varepsilon h_a(x,u) = \mathcal{D}^\star_a \varepsilon(x,u)
\ee
where by $\mathcal{D}^\star_a$ we denote the MHS covariant derivative acting on the MHS tensors,
\be
\mathcal{D}^\star_a \equiv \partial^x_a + i [ h_a(x,u) \stackrel{\star}{,} \quad ] \;.
\ee 

Using the covariant exterior derivative we define the MHS field strength in the standard way
\be \label{Gdef}
\mathbf{F} = \mathbf{D}^\star \mathbf{h} = \mathbf{d} \mathbf{h} + i\, \mathbf{h} \,\hat{\star}\, \mathbf{h}
\ee
which in our case gives
\be \label{omhps}
F_{ab}(x,u) = \partial^x_a h_b(x,u) - \partial^x_b h_a(x,u) + i\, [ h_a(x,u) \stackrel{\star}{,} h_b(x,u)] \;.
\ee
By construction, the MHS field strength transforms covariantly, in the adjoint representation
\be
\mathbf{F}^g(x) = g(x) \mathbf{F}(x) g(x)^{-1}
\ee
which in our case of MHS transformations is
\be \label{Gtent}
F^\mathcal{E}_{ab}(x,u) = e_\star^{-i\, \mathcal{E}(x,u)} \star F_{ab}(x,u) \star e_\star^{i\, \mathcal{E}(x,u)} \;.
\ee
For infinitesimal MHS variations this becomes
\be
\delta_\varepsilon F_{ab}(x,u) = i\, [F_{ab}(x,u) \stackrel{\star}{,} \varepsilon(x,u)] \;.
\ee
It is straightforward to show that the MHS field strength obeys the Bianchi identity
\be \label{omricc}
\mathbf{D}^\star\, \mathbf{F} = 0 \;.
\ee

In YM theory field strength measures the triviality of the configuration. The same applies here, which means that the MHS potential is trivial on an open subset of $M_d$ if and only if the MHS field strength vanishes there
\be \label{hsctc1}
\mathbf{h}\;\; \mbox{is pure gauge} \qquad \Longleftrightarrow \qquad \mathbf{F} = 0 \;.
\ee
This can be proven in the similar fashion as it is usually done in standard YM theory, using the fact that the Moyal product satisfies the algebraic properties of matrix multiplication. The proof of (\ref{hsctc1}) is presented in appendix \ref{app:omtrivproof}.\footnote{At this point the definition of being pure gauge is that every MHS potential $h_a(x,u)$ satisfying a non-degeneracy condition can be obtained by a gauge transformation from the configuration $h_a=0$. A more complete understanding of the equivalence (\ref{hsctc1}) will be possible after introducing the covariant (or geometrical) formulation in section \ref{sec:hsgeom}. See appendix \ref{app:omtrivproof} and section \ref{ssec:hsYM} for more details.}

Before moving to construction of explicit MHS theories, let us briefly comment on the structure of the rigid MHS symmetry in the gauge sector. From (\ref{hsvha}) follows
\be \label{rvmhsp}
\delta_\varepsilon h_a(x,u) = 2\, \varepsilon(u) \sin \left( \frac{1}{2} \stackrel{\leftarrow}{\partial}_u \!\cdot\!
 \stackrel{\rightarrow}{\partial}_x \right) h_a(x,u) \;.
\ee
If we expand the MHS parameter as in (\ref{gHsse}) one realizes that the MHS potential (as any MHS tensor) in general transforms differently than the spacetime matter fields whose transformation is given in (\ref{rhss}). The only exceptions are low-spin transformations with $s\le2$.\footnote{MHS potential is real and so is neutral under spin-1 $U(1)$ transformations, so the only non-trivial case is $n=2$.} Let us demonstrate this on the case of a pure spin-3 variation $\varepsilon(u) = \zeta^{\mu\nu} u_\mu u_\nu$ for which (\ref{rvmhsp}) gives
\be
\delta^{(3)}_\zeta h_a(x,u) = u_\mu\, \zeta^{\mu\nu} \partial_\nu^x h_a(x,u) \;.
\ee
A real spacetime matter field, in comparison, is neutral under the rigid spin-3 variation.

A similar departure from the standard form for rigid HS transformations, given in (\ref{rhss}), appears for matter described by elementary master fields. This is a price to be paid for a simple and consistent formalism. It is also one of the places on which we deviate from standard Noether constructions based on HS symmetries.

\subsection{MHS Yang-Mills model}
\label{ssec:YMa}

Before we continue with the formal mathematical development of the idea, let us examine the physical aspects encoded in it. Following the standard YM construction the natural candidate for an action in the MHS gauge sector is the MHS Yang-Mills (MHSYM) action defined by
\be \label{shsym}
S_{\mathrm{ym}} &=& - \frac{1}{4 g_{\mathrm{ym}}^2} \int d^dx\, d^du\, F^{ab}(x,u) \star F_{ab}(x,u)
\nonumber \\
&=& - \frac{1}{4 g_{\mathrm{ym}}^2} \int d^dx\, d^du\, F^{ab}(x,u) F_{ab}(x,u) + \textrm{(boundary terms)} \;.
\ee
Frame (Latin) indices are contracted using the Minkowski metric $\eta_{ab}$.\footnote{We use the convention  
$\eta_{ab} = \mathrm{diag}(-1,1,\ldots,1)$.} In many applications, such as finding equations of motion (EoM) or analyzing proper gauge symmetries, boundary terms in the action can be discarded. However, there are instances in which one has to be more careful, e.g., analyses of asymptotic symmetries and non-trivial topological structures. The action is weakly non-local, but has at most quartic interaction terms. EoM are
\be \label{ymeom}
\Box_x h_a - \partial^x_a \partial^x_b h^b 
 + i \left( 2 [h^b \stackrel{\star}{,} \partial^x_b h_a] - [h_b \stackrel{\star}{,} \partial^x_a h^b]
 + [\partial^x_b h^b \stackrel{\star}{,} h_a]  \right)
 + \big[ h^b \stackrel{\star}{,} [h_a \stackrel{\star}{,} h_b] \big] = 0 \,. \qquad
\ee

The MHSYM theory was proposed in this form in \cite{paper2}. Note that it is based on the simplest symmetry group for the spin-1 sector, which is $U(1)$. Had we started by assuming a non-abelian symmetry group $G$ (e.g., $SU(N)$), the MHS construction would have been naturally generalized by lumping together the "standard" internal YM structure with the MHS structure \cite{paper2}. In this case the MHS potential would be valued in the Lie algebra $\mathfrak{g}$ of the group $G$. To keep the presentation as simple as possible we shall constrain ourselves in this paper to the $U(1)$ spin-1 group. 

In section \ref{sec:hsgeom} we shall generalise the YM-like formalism which will allow us to see that this form describes just one phase of the theory. Before that, let us mention some of its basic properties.

\begin{itemize}
\item
\emph{Noncommutative structure.}
The MHSYM model generally falls into the class of non-commu\-tative (NC) theories in the broad sense. However, it differs from the standardly studied NC-YM theories. In particular, here the spacetime by itself is commutative and non-commutativity is present only between the spacetime and the auxiliary space.\footnote{In \cite{deMedeiros:2004wb} a somewhat similar model was constructed by taking the associative limit of a gauge theory defined on non-associative fuzzy spaces, the difference being that in \cite{deMedeiros:2004wb} the star product which does not satisfy the hermicity property (\ref{ccmoyp}) appeared instead of the Moyal product.} It is however possible to interpret MHSYM model as a singular limit of a particular NC-YM theory in $2d$ dimensions. It is known that NC-YM field theory defined in a background with a constant metric tensor $G_{ij}$ can be rewritten as\footnote{For a review of NC field theory see \cite{Douglas:2001ba}.}
\be 
S_{\mathrm{nc}} &=& - \frac{1}{4 g_{\mathrm{nc}}^2} \int d^{2d} \widehat{x}\, \sqrt{G}\, G^{ik}G^{jl}{F}_{ij}(\widehat{x}) \star {F}_{kl}(\widehat{x})\,,
\label{MoyalAction}
\ee
where $\widehat{x}=\{\widehat{x}^i \}$ are coordinates and $i,j,\ldots$ are indices in the $2d$-dimensional space. NC gauge field strength is
\be \label{omhps1}
F_{ij}(\widehat{x}) = \partial_i h_j(\widehat{x}) - \partial_j h_i(\widehat{x}) + i\, [ h_i(\widehat{x}) \stackrel{\star}{,} h_j(\widehat{x})]\,.
\ee
The star product encodes NC structure and can be written as
\be
a(\widehat{x}) \star b(\widehat{x}) =
\left. e^{\frac{i}{2}\theta^{ij}\frac{\partial}{\partial\xi^i}\frac{\partial}{\partial\zeta^j}} a(\widehat{x}+\widehat{\xi}) b(\widehat{x}+\widehat{\zeta})\right\vert_{\widehat{\xi}=\widehat{\zeta}=0} \;,
\label{eq:MoyalStar1}
\ee
where $\theta^{ij}$ is a constant antisymmetric matrix. The action is invariant under gauge transformations
\be \label{hsvha1}
\delta_\varepsilon h_i(\widehat{x}) = \partial_i \varepsilon(\widehat{x}) + i\, [h_i(\widehat{x}) \stackrel{\star}{,} \varepsilon(\widehat{x})]\,. 
\ee
Let us now consider a special case in which $G$ and $\theta$ are given by
\be \label{ncymG}
G\equiv 
\begin{pmatrix}
\eta & 0\\
0 & \epsilon^2\, \eta
\end{pmatrix}
\qquad,\qquad
\theta \equiv
\begin{pmatrix}
0 & \eta\\
- \eta & 0
\end{pmatrix} 
\ee
where $\eta$ is $d\times d$ Minkowski metric tensor. If we write $\widehat{x}=\left(x,u\right)$ we see that the star product defined in this way is identical to the Moyal product (\ref{Moyalpr}). Defining
\be
g_{\mathrm{ym}}^2 \equiv g_{\mathrm{nc}}^2\, \epsilon^d
\ee
we obtain
\be
S_{\mathrm{nc}} = S_{\mathrm{ym}} + \mathcal{O}(\epsilon^2) \;.
\ee
In other words, by taking the scaling limit $\epsilon \to 0$ while keeping $g_{\mathrm{ym}}$ fixed and finite, the NC field theory defined by (\ref{MoyalAction})-(\ref{ncymG}) becomes formally identical to the MHSYM action (\ref{shsym}). In this limit $g_{\mathrm{nc}} \to \infty$, so it is a strong coupling limit from the NC-YM field theory perspective. Note that the limit is singular from the perspective of NC field theory because the background metric tensor (\ref{ncymG}) is singular when $\epsilon = 0$.
\item
\emph{Symmetries.}
The MHSYM action is off-shell invariant under the MHS transformations up to boundary terms. This will be important in the study of conservation laws and corresponding conserved charges. There is also a symmetry
\be \label{stsym}
h'_a(x',u') = \Lambda_a{}^b\, h_b(x,u) \quad,\quad x'^\mu = \Lambda^\mu{}_\nu\, x^\nu + \xi^\mu \quad,\quad u'_\mu = \Lambda_\mu{}^\nu\, u_\nu + \tau_\mu
\ee
where $\Lambda$ are Lorentz matrices and $\xi^\mu$ and $\tau_\mu$ are arbitrary constant vectors. The Lorentz transformations acting in spacetime and the auxiliary space must be the same in order to keep the Moyal product invariant. We see that besides the standard Poincar\'{e} group $(\Lambda,\xi)$ of spacetime isometries, there is also an independent group of translations in the auxiliary space. Its existence is a consequence of the fact that the MHSYM action is only weakly non-local on the master space. We show in section \ref{sec:mattHS} that a matter sector may break the auxiliary space translation symmetry.
\item
\emph{Restricted MHS symmetry (even-spin only).} 
The MHSYM action is also symmetric under the auxiliary space reflection
\be
u'_\mu = - u_\mu \qquad,\qquad h'_a(x,u') = - h_a(x,u)
\ee
since the HS field strength transforms as
\be
F'_{ab}(x,u') = - F_{ab}(x,u) \;.
\ee
The MHS variation of the transformed MHS potential keeps the form (\ref{hsvha}), with
\be
\varepsilon'(x,u') = - \varepsilon(x,u) \;.
\ee
It follows that the MHS potential can consistently be restricted to be odd in the auxiliary space 
\be \label{seHSp}
h_a(x,-u) = - h_a(x,u) \;.
\ee
Indeed, it is easy to show that HS transformations are compatible with this if the MHS parameter is also restricted to be odd
\be \label{seHSs}
\varepsilon(x,-u) = - \varepsilon(x,u) \;.
\ee
Using the expansion (\ref{gHsse}) it follows that this restriction corresponds to gauging only spin-even rigid HS transformations.
\item
\emph{Coupling constant.}
The dimension of the MHS potential is (length)$^{-1}$ so the dimension of the coupling constant $g_{\mathrm{ym}}$ is (length)$^{-2}$ for all $d$. It appears that the MHSYM theory has a scale, which we denote by $\ell_h$,  already at the classical level. However, we shall show in section \ref{ssec:hsYM} that the theory in the classical regime does not have an intrinsic scale and that the scale symmetry is spontaneously broken by the choice of the vacuum. To make contact with the canonical formalism it is natural to pass to the dimensionless auxiliary coordinates $\bar{u}$ and a rescaled coupling constant $\bar{g}_h$ and the MHS potential $\bar{h}_a$, defined by
\be \label{cannorm}
\bar{u} = \ell_h u \qquad,\qquad  \bar{g}_\mathrm{ym} = \ell_h^{d/2} g_\mathrm{ym}
 \qquad,\qquad \bar{h}_a = h_a/\bar{g}_\mathrm{ym} \;.
\ee
The dimension of $\bar{g}_\mathrm{ym}$ is (length)$^{\frac{d}{2}-2}$, the same as in the standard Maxwell or Yang-Mills theories, and in $d=4$ it is zero. In the canonical normalization cubic terms and quartic terms in the action have the coupling given by $\bar{g}_\mathrm{ym} \ell_h^{D-1}$ and $\bar{g}_\mathrm{ym}^2 \ell_h^D$, respectively, where $D$ is the total number of spacetime derivatives in a given monomial.
\item
\emph{Stability of the vacuum.}
The configuration $h_a(x,u) = 0$ is obviously a solution of EoM. In the weak-field regime, in which one can ignore terms in the action higher then quadratic, the action is formally similar to the one for the Maxwell theory, so we can immediately obtain the expression for the spatial energy density
\be \label{emtlin}
U \approx \frac{1}{2 g_{\mathrm{ym}}^2} \int d^{d-1}\mathbf{x} \int d^du
 \bigg( \sum_j F_{0j}(x,u)^2 + \sum_{j<k} F_{jk}(x,u)^2 \bigg)
\ee
which is manifestly positive definite and vanishes only for $h_a(x,u) = 0$ (and gauge related configurations). We see that $h_a(x,u) = 0$ is a perturbatively stable vacuum in the classical MHSYM theory. There are no runaway modes in the linearised theory, though one cannot exclude the possibility of their existence in the full theory, due to its non-locality and non-linearity. Note again that we have to impose on the MHS potential proper fall-off conditions at the boundary (infinity) of the auxiliary space to make sure that the action and observables such as energy and momentum are well-defined.
\item
\emph{$L_\infty$ structure.}
All theories based on the MHS symmetry, including the MHSYM theory, provide a representation of the $L_\infty$-algebra \cite{BCDGPS,paper1}.
\item
\emph{Quantization.}
The MHSYM theory can be formally quantized using the BRST method \cite{paper2}.
\item
\emph{Supersymmetry.} The theory can be supersymmetrized \cite{Bonora:2020aqp}.
\end{itemize}

We return to the MHSYM theory and its generalizations in section \ref{sec:mbflt}, and its coupling to matter in section \ref{sec:mattHS}.

\subsection{Physical content and the spacetime description}
\label{ssec:physcon}

In the MHS formalism a natural description is in terms of fields defined on the master space. The classical configuration space is restricted by the requirement that master fields have well-defined Moyal products. Furthermore, due to the infinite volume of the auxiliary space the MHS potential must satisfy proper fall-off conditions, so as to guarantee convergence of integrals over auxiliary space in the expressions for observables such as energy and momentum (see, e.g., (\ref{emtlin})).

The natural objective would be to understand the content of the theory in terms of Wigner's classification of irreducible representations of the Poincar\'{e} group. In particular, is there a purely spacetime description in terms of the standard (finite spin) spacetime fields? Now, by looking at the Eq. (\ref{gHsse}) the simplest assumption would be to assume that such a description is provided by the Taylor expansion in the auxiliary space\footnote{Using an auxiliary space to formally pack a tower of HS spacetime fields into some generator master field, as in (\ref{hscte}), is a standard trick used in conventional HS constructions and analyses of EoM's.} 
\be \label{hscte}
h_a(x,u) = \sum_{n=0}^\infty h_a^{(n)\mu_1\cdots\mu_n}(x)\, u_{\mu_1} \ldots u_{\mu_n} \;.
\ee
The coefficients in the expansion are spacetime fields that are Lorentz tensors of rank $n+1$ symmetric in their $n$ (Greek) indices, and which by (\ref{ymeom}) satisfy EoM that are of the form 
\be \label{ymsteom}
\Box h_a^{(n)\mu_1\cdots\mu_n} - \partial_a \partial^b h_b^{(n)\mu_1\cdots\mu_n} + O(h^2) = 0 
\ee
We see that the linearized EoM for spacetime fields defined by (\ref{hscte}) have Maxwell form with respect to frame indices, but due to the special role of the latter (which bears no symmetry relations with spacetime indices) they are not of the type usually mentioned in the literature (it is neither Fronsdal nor Maxwell as usually defined). A totally symmetric Lorentz tensor field of rank $n$ satisfying Maxwell-like EoM contains irreducible representations with spins $s = n, n-2,n-4,\ldots,1$ or 0, see \cite{Campoleoni:2012th}. As the spacetime fields defined by (\ref{hscte}) have one frame index which is not symmetrized in any way with other (greek) indices, they presumably propagate additional irreducible Poincare representations with spins $s\le n+1$. In spite of this, for shortness' sake, we shall refer to the spacetime field $h_a^{(n)}$ as the spin-$(n+1)$ field.

The expansion (\ref{hscte}) leads us to the spacetime description in terms of infinite tower(s) of HS spacetime fields with unbounded spin. If we restrict the HS potential to an odd function in the auxiliary space (\ref{seHSp}) the tower will only contain spacetime fields with even spin. In \cite{BCDGPS,paper2} it was shown that HS spacetime fields defined by (\ref{hscte}) linearly couple to the corresponding HS currents when spacetime matter fields are minimally coupled to the MHS potential (see section \ref{ssec:framelth} and in particular Eq.\ (\ref{mDfc})). Also, it is straightforward to show that the truncation to $n=0$ and $n=1$ sectors is consistent both with the HS transformations and the MHSYM EoM, meaning that on the level of EoM the low-spin sectors ($s=1$ or 2) may be decoupled from the true HS sector ($s>2$). These low-spin truncations are analyzed in more detail in section \ref{sec:lows}. 

From the form of the HS equations (\ref{ymsteom}) one could be tempted to conclude that the theory has ghosts and, as a consequence, that unitarity is violated. However, such a conclusion is out of reach, since the expansion (\ref{hscte}) which led to the linearised EoM (\ref{ymsteom}) is not suited for the purpose of obtaining a purely spacetime off-shell description. If we substitute (\ref{hscte}) in the MHSYM action and group the terms by order of $u_\mu$, integrations over the auxiliary space would be divergent at each separate order. As we mentioned earlier, proper fall-off conditions in the auxiliary space are required, and the expansion (\ref{hscte}) with {\it infinitely many} terms leads to no problems when all the terms are taken together in calculations. Nevertheless, considering any particular term within the sum on its own, be it in the EoM, action or observables, is in general not allowed. A consequence is that we cannot obtain regular expressions for classical observables, such as energy and momentum, if only a finite number of spacetime fields $h_a^{(n)\mu_1\ldots\mu_n}(x)$ are non-vanishing, in particular  low-spin ($s\le2$) on-shell truncation mentioned above is illusory. The problem with expansion (\ref{hscte}) is not that it is mathematically incorrect or completely useless, but that the spacetime fields defined by it cannot be treated as independent. If one insists on an expansion like (\ref{hscte}), the existence of "hair" consisting of an infinite tail of HS spacetime fields is obligatory in physically acceptable configurations. The main conclusion is that {\it spacetime fields generated by the expansion} (\ref{hscte}) {\it do not reflect the spectrum (particle content) of the MHSYM theory}. 

No expansion in a discrete basis of functions in the auxiliary space exists which provides at the same time (i) convergence of integrals over auxiliary space inside the action needed for the purely spacetime off-shell description, and (ii) manifest Lorentz covariance. One may try to bypass the requirement (i) and regularize integrals over master space by absorbing the (infinite) volume through a (classical) renormalization of the coupling constant. Such an approach was proposed in \cite{paper2} with partial success. The problem with this strategy in general is that the theory is essentially redefined, which introduces the danger of losing some of the crucial qualities present in the classically regular master space description, such as perturbative stability of the vacuum. 

Alternatively, one can ignore the requirement (ii) and consider expansions in an orthonormal discrete basis of functions  $\{f_r(u)\}$,
\be \label{mpone}
h_a(x,u) = \sum_r h_a^{(r)}(x)\, f_r(u)
\ee
where 
\be \label{fuonb}
\int d^d u\, f_r(u)\, f_s(u) = \delta_{rs} \;.
\ee
Such an expansion necessarily breaks the full manifest Lorentz covariance in the $(x,u)$ (master space), in the sense that the spacetime fields $h_a^{(r)}(x)$ are labeled by the index ($r$) which is not one of a covariant Lorentz representations.\footnote{In this approach it is implicitly assumed that there exists a discrete basis which leads to a unitary spacetime description.} Using such an expansion one arrives at the off-shell space time description with the quadratic part of the Lagrangian given by
\be
S_0[h] = - \frac{1}{4 g_{\mathrm{ym}}^2} \sum_r \int d^dx\, \big(\partial_a h_b^{(r)} - \partial_b h_a^{(r)}\big)
\eta^{ac} \eta^{bd} \delta_{rs} \big(\partial_c h_d^{(s)} - \partial_d h_c^{(s)}\big) \;.
\ee
The above linearised action neither contains dangerous ghosts (of the kind that cannot be removed using gauge freedom), nor runaway modes. In a similar way one can integrate the interacting part of the MHSYM action over the auxiliary space to obtain a purely spacetime action which is a weakly non-local functional of spacetime fields $\{ h_a^{(r)}(x) \}$.

In summary, we conclude that {\it the MHSYM theory apparently does not have a Lagrangian (off-shell) description in terms of (a tower of) HS spacetime fields carrying standard massless finite spin irreducible representations of the Poincar\'{e} group}. On the one hand this may seem strange considering our starting point was a tower of global HS symmetries, but on the other hand it is not surprising considering that attempts based on standard HS fields in Minkowski spacetime background not only failed but also produced a number of no-go theorems. 
While to prove unitarity and ultimately a consistency of the MHSYM theory one would need to study full interacting quantized theory, an effort which goes beyond the scope of this paper, we have shown that spacetime formulations in which manifest Lorentz covariance is broken are free of dangerous ghosts. It remains to be seen whether manifest Lorentz covariance may be eventually reconstructed, or the boundary conditions in the auxiliary space introduce some sort of the soft breaking of Lorentz symmetry. We leave this question for our future work.

\subsection{Euclidean theory}
\label{ssec:euclid}

It is interesting that a viable purely spacetime off-shell description is possible in the Euclidean version of the theory. In this case it is easy to find regular expansions of the HS master potential leading to convergent integrals over the auxiliary space, which are manifestly covariant under the isometries of the flat Euclidean spacetime.  

The simplest choice closely resembling (\ref{hscte}) would be
\be \label{regute}
h_a(x,u) = \sum_{n=0}^\infty \tilde{h}_a^{(n)\mu_1\cdots\mu_n}(x)\, u_{\mu_1} \ldots u_{\mu_n}\, e^{-(\ell_h u)^2/2}
\ee
where $u^2 = \delta^{\mu\nu} u_\mu u_\nu$. Comparing with (\ref{hscte}) we see that the difference is the exponential factor. This factor guarantees convergence of all integrations over auxiliary space after using (\ref{regute}) in the MHSYM action or in the energy-momentum spacetime tensor. In this way we obtain a regular Euclidean spacetime action and other observables as functionals of the infinite tower of spacetime potentials $\{\tilde{h}_a^{(n)\mu_1\cdots\mu_n}(x)\}$. A downside of this approach is the lack of a consistent on-shell truncation to a finite subset of spacetime fields.

Let us apply (\ref{regute}) to the quadratic (free field) part of the MHSYM action in the canonical normalization (\ref{cannorm})
\be \label{hsym2}
S_{\mathrm{ym}}^{(0)}[\tilde{h}] = \frac{1}{4 g^2_\mathrm{ym}} \int d^dx \int d^du\, 
\big(\partial_a \tilde{h}_b(x,u) - \partial_b \tilde{h}_a(x,u)\big)^2 \;.
\ee
Integrating over the auxiliary space we obtain the quadratic part of the action in the (purely) spacetime description
\be
S_{\mathrm{ym}}^{(0)}[\tilde{h}] = \int d^dx\, \mathscr{L}_0(x)
\ee
with the spacetime Lagrangian density given by
\be \label{hsymeq}
\mathscr{L}_0 = \sum_{\substack{n,m=0 \\ n+m \mathrm{\ even}}}^\infty \!\!\!
 \big( \partial_a \tilde{h}_b^{(n)\mu_1\ldots\mu_n} - \partial_b \tilde{h}_a^{(n)\mu_1\ldots\mu_n} \big)
 \mathcal{K}^{(n+m)}_{\mu_1\ldots\mu_n\nu_1\ldots\nu_m} \big( \partial_a \tilde{h}_b^{(n)\nu_1\ldots\nu_m}
 - \partial_b \tilde{h}_a^{(n)\nu_1\ldots\nu_m} \big) . \qquad
\ee
The kinetic matrix is totally symmetric and given by
\be \label{hsymept}
\mathcal{K}^{(2r)}_{\mu_1\ldots\mu_{2r}} = c_d\, \frac{\ell_h^{-(d+2r)}}{4\, g_{\mathrm{ym}}^2}\, 
\Gamma\Big(r + \frac{d}{2}\Big)\, \frac{(2r-1)!!}{(d+2r-2)!!}\, \delta_{(\mu_1\mu_2} \cdots \delta_{\mu_{2r-1}\mu_{2r})} 
\ee
where $c_d$ is a (finite) numerical coefficient depending on $d$ (but not on $m$ or $n$) whose exact expression is not of relevance here. We see that the quadratic term is not diagonal in the spacetime HS fields, but strongly mixed. In fact, it is as non-diagonal as it gets, because rotation invariance prevents mixing between spin-odd (even $n$) and spin even (odd $n$) fields in the quadratic Lagrangian. This complicates the extraction of the physical spectrum and calculation of the propagator, as for this one first has to invert the kinetic matrix (\ref{hsymept}).\footnote{Such mixing of HS fields in the linearised EoM was observed in \cite{Bonora:2017ykb} by using induced action methods.}

As in the Lorentzian case, the problem of diagonalizing the quadratic part of the action can be avoided using an orthonormal basis of functions in the auxiliary space $\{f_r(u)\}$ with (\ref{mpone})-(\ref{fuonb}). The novelty here is that in the Euclid space one can easily find orthonormal bases which are manifestly covariant under $SO(d)$ rotations. One choice would be to use spherical harmonics. The other, which has the benefit of having spacetime fields being represented by standard $SO(d)$ tensors (as in (\ref{regute})), is obtained from Hermite functions. In this way one obtains purely spacetime description in terms of infinite tower of fields 
$\{ h_a^{(r)}(x) \}$, with the action having the properties: (i) manifest covariance under (Euclid) spacetime symmetries, (ii) semi-positive definiteness, (iii) quadratic part is first order in derivatives and is given by
\be \label{Msa0}
S_0[h] = - \frac{1}{4 g_{\mathrm{ym}}^2} \sum_r \int d^dx\, \big(\partial_a h_b^{(r)} - \partial_b h_a^{(r)}\big)
\delta^{ac} \delta^{bd} \delta_{rs} \big(\partial_c h_d^{(s)} - \partial_d h_c^{(s)}\big) \;,
\ee
(iv) interacting part is a weakly non-local functional of spacetime fields. 

Note that sets (infinite towers in fact spacetime fields defined with respect to different bases in the auxiliary space mentioned above, are connected by field redefinitions of the algebraic type so in principle one can pass from one to the other. The expansions based on standard tensors, such as (\ref{regute}) or the one based on Hermite functions, has the advantage that the spacetime fields carry representations of both Euclid and Minkowski isometries ($SO(d)$ and $SO(1,d-1)$, respectively), which is beneficial from the perspective of the Wick rotation (if such a rotation is meaningful in MHS theories defined on the master space).

\section{MHS formalism in the covariant formulation}
\label{sec:hsgeom}

\subsection{MHS tensors}

Here we explore the MHS structure introduced in section \ref{ssec:mhsstr} in more depth. We use a formal similarity with non-commutative field theories to borrow some of the techniques, and show that there exists a covariant frame-like formulation. This formulation will be important in understanding the emergent geometrical description of the spin-2 sector. As we will show, it also offers a better starting point for a background independent formulation in terms of matrix models.

In the YM-like approach the basic object that covariantly transforms, in the adjoint representation, under MHS transformations is the MHS master field strength, see (\ref{Gtent}). Any master field $A(x,u)$ transforming in the same way, which is
\be \label{hslarge}
A^\mathcal{E}_{ab\cdots}(x,u) = e_\star^{-i\, \mathcal{E}(x,u)} \star A_{ab\cdots}(x,u) \star
 e_\star^{i\, \mathcal{E}(x,u)}
\ee
we call an MHS tensor. An MHS tensor in general can have any number of frame-like indices (denoted by Latin letters $a,b,\ldots$), on which MHS transformations do not act. For the moment we assume flat background and trivial frames, so frame indices are raised and lowered with the Minkowski metric tensor. For infinitesimal MHS transformations this gives the MHS variation
\be \label{hsvar}
\delta_{\varepsilon} A_{a\cdots}(x,u) = i\, [ A_{a\cdots}(x,u) \stackrel{\star}{,} \varepsilon(x,u) ] \;.
\ee
The important property of MHS tensors is that the Moyal product of MHS tensors is again an MHS tensor. This is a trivial consequence of (\ref{hslarge}) and (\ref{hsinv}).

\subsection{MHS vielbein}

In standard YM gauge theories, to construct a covariant object from the gauge potential we need to take a derivative. This object is the gauge field strength. Non-commutativity of the MHS structure allows us to construct an MHS tensor without using derivatives, in the following way
\be \label{gvieldef}
e_a(x,u) \equiv u_a + h_a(x,u) \;.
\ee
Using (\ref{hsvha}) it is easy to show that $e_a(x,u)$ transforms under MHS variations as
\be \label{esvha}
\delta_\varepsilon e_a(x,u) = i\, [ e_a(x,u) \stackrel{\star}{,}  \varepsilon(x,u) ] \;.
\ee
which is exactly (\ref{hsvar}). The presence of such an object is not unexpected from the viewpoint of NC field theories. By using it instead of the MHS potential $h_a(x,u)$ we can write all equations in the MHS gauge sector in a manifestly MHS covariant way (i.e., by using exclusively MHS tensors), a feat not possible in the standard YM theories. 

We refer to $e_a(x,u)$ as the MHS vielbein.\footnote{Note that $e_a(x,u)$ can be understood as MHS covariant coordinates on the auxiliary space.} The motivation comes from analysing its Taylor expansion in the auxiliary coordinates
\be \label{hsvte}
e_a(x,u) = \sum_{n=0}^\infty e_a^{(n)\mu_1\ldots\mu_n}(x)\, u_{\mu_1} \cdots u_{\mu_n} \;.
\ee
As we show in section \ref{sec:lows} the spin-2 ($n=1$) spacetime component transforms under the spin-2 part of the MHS transformations as a vector frame under diffeomorphisms. Moreover, if we assume (\ref{seHSp}), which is consistent with the MHS symmetry, this component is the lowest term in the expansion. When coupled to spacetime matter, this vector frame plays the role of the vielbein, as shown in section \ref{ssec:framelth}.

This expansion also illuminates the meaning of (\ref{gvieldef}). Performing Taylor expansions (\ref{hsvte}) and (\ref{hscte}) one obtains that the corresponding spacetime fields are connected through 
\be
e_a^{(n)\mu_1\ldots\mu_n}(x) = h_a^{(n)\mu_1\ldots\mu_n}(x) \qquad,\qquad n\ne1
\ee
and
\be \label{hsvs2}
e_a^{(1)\mu}(x) = \delta_a{}^\mu + h_a^{(1)\mu}(x) \;.
\ee
We see that (\ref{gvieldef}) defines the MHS potential with respect to the empty Minkowski background
\be
e_a = u_a \equiv \delta_a^\mu u_\mu \;.
\ee
This is not surprising, but it shows the limits of practical usability of (\ref{gvieldef}). We can again see that the MHS vielbein is the fundamental object in the theory, and that (\ref{gvieldef}) is sensible only if we are interested in expansions around the empty Minkowski vacuum. 

Strictly speaking, to identify $e_a^{(1)\mu}(x)$ as a spacetime vielbein, an invertibility condition should be imposed. This condition is apparently not required in the MHS formalism, which opens the possibility of accommodating configurations and phases with non-geometric interpretations.

\subsection{MHS covariant derivative and torsion}
\label{ssec:HScd}

In section \ref{ssec:mhsstr} we have noted that 
\be \label{hscd}
\mathcal{D}^\star_a = \partial^x_a + i\, [h_a(x,u) \stackrel{\star}{,} \quad ]
\ee
fulfills al the requirements usually requested from a covariant derivative when acted on an MHS tensor.\footnote{The requirements are that (i) it is gradient linear, (ii) it maps tensors into tensors, (iii) obeys the Leibniz rule and (iv) it is the inverse of the integral, which in our context means
\be \label{vanintdiv}
\int d^dx\, d^du\, \mathcal{D}^\star_a\, A^{a\ldots}(x,u) = \textrm{(boundary terms)} \;.
\ee
Using the properties of the Moyal product listed in section \ref{app:moyal} it is easy to show that the definition (\ref{hscdc}) satisfies all four properties.} Using (\ref{gvieldef}) we obtain that the background independent formulation of (\ref{hscd}) is given by
\be \label{hscdc}
\mathcal{D}^\star_a = i\, [e_a(x,u) \stackrel{\star}{,} \quad ] \;.
\ee
As we will see soon, this form is not only more generic but also usually more convenient for performing calculations. Note that we can write the MHS variation of the MHS vielbein in a manifestly covariant form as
\be \label{deacov}
\delta_{\varepsilon} e_a(x,u) = \mathcal{D}^\star_a\, \varepsilon(x,u) \;.
\ee

Having defined the MHS vielbein and covariant derivative a natural object to construct is
\be \label{hsscp}
T_{ab}(x,u) \equiv \mathcal{D}^\star_a\, e_b(x,u)
 = i\, [e_a(x,u) \stackrel{\star}{,} e_b(x,u)]
\ee
which is an antisymmetric MHS tensor
\be
T_{ab}(x,u) = - T_{ba}(x,u) \;.
\ee
As we show in section \ref{ssec:interpr12}, the Moyal bracket in the spin-2 sector behaves as the Lie bracket of vector fields. It then follows from (\ref{hsscp}) that $T_{ab}$ can be interpreted both as the generalized anholonomy and the generalized torsion.\footnote{The latter is obvious when we write $T_{ab}$ in the following form
\be
T_{ab} = \mathcal{D}^\star_a\, e_b(x,u) - \mathcal{D}^\star_b\, e_a(x,u) - i\, [e_a(x,u) \stackrel{\star}{,} e_b(x,u)] \;.
\ee} 
We shall refer to it as the MHS torsion. Expanding around flat background (\ref{gvieldef}), we get
\be \label{omhst}
T_{ab}(x,u) = \partial^x_a h_b(x,u) - \partial^x_b h_a(x,u) + i\, [ h_a(x,u) \stackrel{\star}{,} h_b(x,u)]
\ee
which is the MHS field strength obtained in the YM-like construction and defined in (\ref{omhps}). Our convention is to use the symbol $T_{ab}$ in generic situations, and the symbol $F_{ab}$ when (\ref{gvieldef}) is meaningful.

One could ask what is the HS generalization of the Riemann tensor. In differential geometry the Riemann tensor is extracted from the commutator of covariant derivatives.  The commutator of MHS covariant derivatives, acting on an arbitrary MHS tensor, gives
\be \label{comHScd}
[ \mathcal{D}^\star_a\, , \mathcal{D}^\star_b ] A_{c\ldots}(x,u)
 = i\, [ T_{ab}(x,u) \stackrel{\star}{,} A_{c\ldots}(x,u) ]
\ee
thus it is defined by the MHS torsion. As a special case,
\be
[ \mathcal{D}^\star_a\, , \mathcal{D}^\star_b ]\, e_c(x,u) = \mathcal{D}^\star_c T_{ba}(x,u)
\ee
from which we see that there is no extra independent structure in our formalism corresponding to the generalized Riemann tensor. Using the Jacobi identity (\ref{mbji}) it is straightforward to show that the MHS torsion satisfies the MHS Bianchi identity, 
\be \label{omricc2}
\mathcal{D}^\star_a\, T_{bc}(x,u) + \mathcal{D}^\star_b\, T_{ca}(x,u) + \mathcal{D}^\star_c\, T_{ab}(x,u) = 0
\ee
which is the same as (\ref{omricc}).

Note that by putting $A_{c\ldots}(x,u) = \varepsilon(x,u)$ in (\ref{comHScd}) we can write the MHS variation of the MHS torsion as
\be \label{omvar}
\delta_\varepsilon T_{ab}(x,u) &=& \left[ \mathcal{D}^\star_a\, , \mathcal{D}^\star_b \right]\, \varepsilon(x,u)
\\
&=& \mathcal{D}^\star_a\, \delta_\varepsilon e_b(x,u) - \mathcal{D}^\star_b\, \delta_\varepsilon e_a(x,u) \;.
\label{deltab}
\ee

\subsection{The MHS metric}
\label{sssec:mhsmet}

The simplest HS tensor without frame indices in our formalism is
\be \label{gvmconn}
g(x,u) \equiv e_a(x,u) \star e^a(x,u) \;.
\ee
For the obvious reason we call it the MHS metric. If we use (\ref{hsvte}) and a similar Taylor expansion for the MHS metric
\be \label{hsmte}
g(x,u) = \sum_{s=0}^\infty g_{(s)}^{\mu_1\ldots\mu_s}(x)\, u_{\mu_1} \cdots u_{\mu_s}
\ee
it follows from (\ref{gvmconn}) that the $s=2$ component is given by
\be \label{g2ee}
g_{(2)}^{\mu\nu}(x) = \eta^{ab} e_a^{(1)\mu}(x)\, e_b^{(1)\nu}(x) +  (\mbox{HS contributions})
\ee
where every monomial in "(HS contributions)" contains field(s) $e_a^{(n)\mu_1\ldots\mu_n}(x)$ with $n\ge2$, which is spin $\ge3$. Up to spin $s>2$ contributions, this is exactly the relation between a metric and a vielbein in standard differential geometry.

If we expand the MHS vielbein as in (\ref{gvieldef}), then the natural way to expand the HS metric is
\be \label{gmetric}
g(x,u) \equiv u^2 + h(x,u) \;.
\ee
Taylor expanding both sides around $u=0$, we get for $s\ne2$
\be
g_{(s)}^{\mu_1\ldots\mu_s}(x) = h_{(s)}^{\mu_1\ldots\mu_s}(x)  \qquad,\qquad s\ne2
\ee
and for $s=2$
\be \label{gms2}
g_{(2)}^{\mu\nu}(x) = \eta^{\mu\nu} + h_{(2)}^{\mu\nu}(x) \;.
\ee
We see that the MHS field $h(x,u)$ measures the deviation from the flat background. Using (\ref{gvmconn}), (\ref{gmetric}) and (\ref{esvha}) we get the MHS variation of $h(x,u)$
\be \label{delh2}
\delta_{\varepsilon} h(x,u) = 2 (u\!\cdot\! \partial_x)\varepsilon (x,u)
 + i\, [ h(x,u) \stackrel{\star}{,} \varepsilon (x,u)] 
\ee
which is exactly the variation found in \cite{Bekaert:2009ud,Bekaert:2010ky} in the analysis of MHS symmetries of the free Klein-Gordon field linearly coupled to the infinite tower of spacetime HS fields. In 
\cite{Bekaert:2009ud,Bekaert:2010ky} it was argued that $h(x,u)$ should be a composite field, and here we made it explicit. In section \ref{ssec:framelth} we show that $h(x,u)$ indeed is the field which couples minimally to the Klein-Gordon  field in the MHS formalism. Using (\ref{gvmconn}), (\ref{gmetric}) and (\ref{gvieldef}) we obtain
\be \label{hharel}
h(x,u) = 2\, u^a h_a(x,u) + h_a(x,u) \star h^a(x,u) \;.
\ee
In particular the $s=0$ component of $h(x,u)$, which provides seagull vertices for $s\ge1$ interactions when coupled to a Klein-Gordon field, is 
\be
h_{(0)}(x) = h_a(x,u) \star h^a(x,u) \Big|_{u=0} \;.
\ee

The MHS covariant derivative is not metric-compatible since
\be \label{hsnmet}
Q_a(x,u) \equiv \mathcal{D}_a^\star g(x,u) = i\, [ e_a(x,u) \stackrel{\star}{,} g(x,u)]
\ee
is generally not vanishing. We refer to the HS tensor $Q_a(x,u)$ as the MHS nonmetricity tensor. The underlying geometry in our construction appears not to be of the Riemann-Cartan type. Note that the MHS nonmetricity tensor (\ref{hsnmet}) can be written as
\be
Q_a(x,u) = \{ e_b(x,u) \stackrel{\star}{,} T_a{}^b(x,u) \}
\ee
i.e., it is completely determined by the MHS torsion.

To summarize, the geometry emerging in the MHS theory has all fundamental tensors (torsion, Riemann tensor and nonmetricity) non-vanishing. While the geometry may look exotic at a first glance, it is in fact closely related to the teleparallel geometry. In section \ref{sec:lows} we study in detail the spin-2 sector, and show that the emergent linear connection is opposite\footnote{To use the terminology of \cite{GoM}.} to the Weitzenb\"{o}ck connection.

\subsection{Local Lorentz covariance}
\label{ssec:lli}

In the construction above we have drawn analogies with differential geometry. It is then natural to ask can the MHS formalism accommodate the covariance under local Lorentz transformations of the vielbein
\be
E_a{}^\mu(x) \to \Lambda_a{}^b(x)\, E_b{}^\mu(x) \;.
\ee 
If we try to generalize this to the MHS vielbein
\be
e_a(x,u) \to \Lambda_a{}^b(x)\, e_b(x,u)
\ee
we obtain that none of the MHS objects we constructed (metric, torsion, MHSYM action) transform covariantly. Covariance is present only for rigid transformations, $\Lambda_a{}^b(x) = \Lambda_a{}^b$. The apparent failure of local Lorentz covariance is a consequence of not taking properly into account the fact that our construction so far was based on a special choice of the inertial frame. Local Lorentz transformations turn inertial frames into non-inertial, so we have to extend our formalism to non-inertial frames. To accommodate for this we have to introduce the spin connection $\mathscr{A}^a{}_{b\mu}(x)$ in the ambient Minkowski spacetime and use the covariant derivative instead of ordinary one 
\be
\partial^x \to \mathscr{D} = \partial^x + \mathscr{A} 
\ee
inside the Moyal products. Note that the spin connection is flat, which means that $\mathscr{A}^a{}_{b\mu} = \Lambda^a{}_c\, \partial_\mu \Lambda_c{}^b$ and
\be
[\mathscr{D}_\mu , \mathscr{D}_\nu] = 0 \;.
\ee
This guarantees that Moyal products of the MHS tensors transform covariantly under local Lorentz transformations, e.g.,
\be
T_{ab}(x,u) \to \Lambda_a{}^c(x) \Lambda_b{}^d(x)\, T_{cd}(x,u)
\ee
from which follows that the MHSYM theory is locally Lorentz covariant when properly formulated. 

Of course, such introduction of gauge symmetry is trivial and does not influence the degrees of freedom present in the theory. It is emphasized here for the sake of the induced geometry interpretation developed in section \ref{sec:lows}.

\section{Model building: MHS gauge sector}
\label{sec:mbflt}

\subsection{General considerations}
\label{ssec:gcons}

Having developed the MHS formalism and having it cast into a more general form, let us turn back to the question of constructing candidates for the theories based on the MHS symmetry. As for degrees of freedom, we expect to have an MHS gauge sector described by the MHS vielbein $e_a(x,u)$, which is an MHS tensor (transforms in the adjoint representation), and a matter sector spanned by a set of matter fields collectively denoted by $\psi$ which can be in different representations of the MHS symmetry (discussed in more detail in section \ref{sec:mattHS}). Correspondingly, the action is a sum of two parts,
\be \label{MHSmod}
S[e,\psi] = S_{\mathrm{hs}}[e] + S_{\mathrm{m}}[\psi,e]
\ee
one describing the pure MHS gauge theory, and the other the matter sector and its coupling to the MHS vielbein.  We assume that the action is weakly non-local in the master space, by which we mean that both parts in (\ref{MHSmod}) can be written in terms of a master space Lagrangian
\be
S_{\mathrm{hs,m}} = \int d^dx\, d^du\, L_{\mathrm{hs,m}}(x,u) \;.
\ee
To keep it simple here we only consider Lagrangian terms which are (Moyal) polynomials manifestly covariant under the MHS symmetry. This means that $L_{\mathrm{hs}}(x,u)$ is an MHS tensor, while $L_\mathrm{m}(x,u)$ may be an MHS scalar or an MHS tensor, depending on the matter content. In both cases MHS invariance of the action
\be \label{mhsagi}
\delta_\varepsilon S[e] = 0
\ee  
is manifestly guaranteed. In case of a master space Lagrangian being an MHS tensor, invariance of the action is a consequence of the trace property of the Moyal product (\ref{mpintp}).\footnote{Assuming proper gauge transformations for which boundary terms vanish.} In addition, we assume that the Lagrangian is a Lorentz scalar.

From (\ref{MHSmod}) it follows that the EoM in the MHS gauge sector is
\be \label{cceom}
0 = \mathcal{F}_\mathrm{hs}^a(x,u) + \mathcal{J}_\mathrm{m}^a(x,u)
\ee
where
\be \label{eomco}
\mathcal{F}_\mathrm{hs}^a(x,u) = \frac{\delta S_\mathrm{hs}[e]}{\delta e_a(x,u)} \qquad,\qquad
\mathcal{J}_{\mathrm{m}}^a(x,u) = \frac{\delta S_\mathrm{m}[\psi,e]}{\delta e_a(x,u)}
\ee
To obtain the EoM, the trace property of the Moyal product can be used, since variations of the fields must vanish at the boundary. This makes it equal whether the functional derivative is "left" or "right". One should in general distinguish between
\begin{equation}
\delta_L A[e,\psi] = \int d^dx\, d^du\, \delta e_a \star \frac{\delta_L A}{\delta e_a},\qquad \delta_R A[e,\psi] = \int d^dx\, d^du\, \frac{\delta_R A}{\delta e_a}\star\delta e_a 
\end{equation}
but on places where it does not make a difference, we will omit an explicit subscript.

The EoM in matter sector are given by
\be \label{gmEoM}
0 = \frac{\delta S_\mathrm{m}[\psi,e]}{\delta \psi} \;.
\ee

Using (\ref{deacov}), (\ref{vanintdiv}) and (\ref{mpintp}) the MHS variation of $S_\mathrm{hs}$ can be written as\footnote{It is assumed here that the MHS variation is a proper gauge transformation, in which case boundary terms in (\ref{vanintdiv}) and (\ref{mpintp}) vanish.}
\be
\delta_\varepsilon S_\mathrm{hs}[e] &=& \int d^dx\, d^du\, \mathcal{F}_\mathrm{hs}^a(x,u) \,  
 \delta_\varepsilon e_a(x,u)
\nonumber \\
&=& \int d^dx\, d^du\, \mathcal{F}_\mathrm{hs}^a(x,u) \, \mathcal{D}^\star_a \varepsilon(x,u)
\nonumber \\
&=& - \int d^dx\, d^du\, \mathcal{D}^\star_a \mathcal{F}_\mathrm{hs}^a(x,u)\, \varepsilon(x,u) \;.
\ee
Then, from (\ref{mhsagi}) we get the off-shell identity
\be \label{hsbianchi}
\mathcal{D}^\star_a \mathcal{F}_\mathrm{hs}^a(x,u) = 0 \;.
\ee
Applying $\mathcal{D}^\star_a$ on EoM (\ref{cceom}) and using (\ref{hsbianchi}) we get
\be \label{hsccons}
\mathcal{D}_a^\star \mathcal{J}_\mathrm{m}^a(x,u) = 0
\ee
which states that the matter master current is (on-shell) covariantly conserved.

\subsection{MHSYM theory}
\label{ssec:hsYM}

Let us first analyze the MHS gauge sector. The matter sector is studied in section \ref{sec:mattHS}. The simplest acceptable Lagrangian term that is dynamical and has the flat vacuum $e_a(x,u) = u_a$ as a solution of the EoM it generates is
\be \label{hsymla}
L_\mathrm{hs}(x,u) = \frac{1}{4 g_\mathrm{ym}^2}\, T_{ab}(x,u) \star T^{ba}(x,u) \equiv L_\mathrm{ym}(x,u) \;.
\ee
This is the Lagrangian of the MHSYM theory already introduced in section \ref{ssec:YMa}. The corresponding action is
\be
S_\mathrm{ym}[e] &=& \frac{1}{4 g_\mathrm{ym}^2} \int d^dx\, d^du\, T_{ab}(x,u) \star T^{ba}(x,u)
\nonumber \\
&=& \frac{1}{4 g_\mathrm{ym}^2} \int d^dx\, d^du\, [ e_a \stackrel{\star}{,} e_b ] \star [ e^a \stackrel{\star}{,} e^b ] \;.
\ee
Under a generic variation $\delta e_a(x,u)$ that vanishes on the boundary of the integration volume, the MHSYM action behaves as
\be
\delta S_\mathrm{ym}[e] &=& \frac{1}{2 g_\mathrm{ym}^2}  \int d^dx\, d^du\, \{ \mathcal{D}^\star_b T^{ba}(x,u) \stackrel{\star}{,} \delta e_a(x,u) \}
\nonumber \\
&=& \frac{1}{g_\mathrm{ym}^2}  \int d^dx\, d^du\, \mathcal{D}^\star_b T^{ba}(x,u)\, \delta e_a(x,u)
\ee
which means that its contribution to the EoM is
\be
\mathcal{F}_\mathrm{ym}^a(x,u) &=& \frac{1}{g_\mathrm{ym}^2}\, \mathcal{D}^\star_b\, T^{ba}(x,u)
\nonumber \\
&=& \frac{1}{g_\mathrm{ym}^2}\, \big[ e_b(x,u) \stackrel{\star}{,} [e^a(x,u) \stackrel{\star}{,} e^b(x,u)] \big] \;.
\label{ymeomt} 
\ee
The EoM of the pure MHSYM theory is then
\be
\mathcal{D}^\star_b\, T^{ba}(x,u) = 0 \;.
\ee
It is important to observe that the MHSYM theory is classically a scale-free theory from the master space perspective. If one takes the MHS vielbein to be dimensionless, then the pure MHSYM coupling constant $g_\mathrm{ym}$ is also dimensionless. Moreover, as the coupling constant can be absorbed by rescaling the MHS vielbein, it is a theory without an intrinsic coupling constant. We have seen that in the YM formulation the theory was not scale-free, so how is this possible? The answer is that the scale was introduced by a choice of the empty flat vacuum. In this normalization, it is given by
\be
e_a(x,u)_\mathrm{vac} = \ell_h u_a \;.
\ee
Note that the scale $\ell_h$ can be changed by "canonical" scale transformations
\be
e_a'(x,u) = e_a(\lambda x, u/\lambda)
\ee
which form a subgroup of MHS transformations. The Minkowski vacuum spontaneously breaks a part of the MHS symmetry.\footnote{There are other mechanisms which might introduce a scale in the MHSYM theory. In section \ref{ssec:euclid} we have seen that this may happen when the theory is written in a purely spacetime form.}

A few words are in order about vacua in MHSYM theory. They are solutions of the EoM which satisfy the condition\footnote{In section \ref{ssec:conslc} we show that all conserved charges, including the energy-momentum tensor, vanish for configurations satisfying (\ref{ymvacc}).}
\be \label{ymvacc}
T_{ab}(x,u) = 0
\ee
We have already mentioned that the flat configuration $e_a(x,u) = u_a$ is, at least from the classical viewpoint, a well-defined Lorentz-invariant vacuum. However, it is not the case that all configurations satisfying (\ref{ymvacc}) are MHS gauge equivalent to the flat vacuum. An obvious example is an "empty" configuration $e_a(x,u) = 0$ which is a fixed point of MHS gauge transformations. To obtain some insight into the structure of vacua, let us examine the vacua that are of the form
\be \label{convac}
e_a(x,u) = M_a{}^\mu u_\mu
\ee
where $M$ are arbitrary constant real $d \times d$ matrices. MHS transformations preserving the shape of these configurations have the gauge parameter master field of the form
\be
\mathcal{E}_\Lambda(x,u) = x^\mu \Lambda_\mu{}^\nu u_\nu
\ee
where $\Lambda$ is again an arbitrary constant real $d \times d$ matrix. Using (\ref{gecvac}) it is easy to show that such an MHS transformation when acting on a vacuum (\ref{convac}) produces the same type of vacuum with matrix $M^\Lambda$ given by
\be
M^\Lambda = M e^\Lambda \;.
\ee
where matrix multiplication is assumed. A corollary is that two vacua of the type (\ref{convac}), which are defined with matrices of different rank, are MHS gauge inequivalent.

This analysis sugests that the MHSYM theory contains different phases. The flat vacuum $e_a(x,u) = u_a$ describes empty flat (Minkowski) background and defines a geometric phase. When expanded around the flat vacuum solution as in (\ref{gvieldef}) the linear part of the EoM is second-order in spacetime derivatives, and in this phase the theory has a perturbative regime (in the coupling constant). In contrast, the configuration $e_a(x,u) = 0$ does not have an emergent regular geometric description and defines a non-perturbative strongly-coupled unbroken phase (it is the only vacuum with a trivial orbit with respect to the MHS transformations).\footnote{By a regular geometric description we mean a description based on the non-degenerate emergent metric tensor or vielbein.}

\subsection{Beyond MHSYM theory}
\label{ssec:genYM}

Let us now analyze possible generalizations of the MHSYM theory by adding additional terms to the action. There is just one lower-dimensional term allowed by the MHS symmetry for a generic number of spacetime dimensions $d$,
\be
L_1(x,u) = \frac{\lambda_1}{2}\, g(x,u) = \frac{\lambda_1}{2}\, e_a(x,u) \star e^a(x,u) 
\ee
which produces the following EoM contribution 
\be
\mathcal{F}_1^a(x,u) = \lambda_1\, e^a(x,u) \;.
\ee
This term is not dynamical and has the appearance of a mass term, but in the geometric phase it behaves as a generalized cosmological constant term. When added to the MHSYM action, the flat configuration $e_a(x,u) = \delta_a^\mu u_\mu$ is no longer solution of EoM, so this term changes the vacuum in the geometric phase.

For general $d$ there is one additional independent term, which is of the same dimension as $F^2$,
\be
L_2(x,u) = \frac{\lambda_2}{4}\, g(x,u) \star g(x,u)
\ee
which would contribute the following EoM term
\be
\mathcal{F}_2^a(x,u) = \frac{\lambda_2}{2}\, \{ g(x,u) \stackrel{\star}{,} e^a(x,u) \} \;.
\ee
Adding this term also removes the flat configuration $e_a(x,u) = u_a$ from the solution space, but there is more. This term is dynamical and so it is interesting to see how it contributes to the linearized EoM in the geometric phase. If we write
\be
e_a(x,u) = e_a^{(0)}(x,u) + h_a(x,u)
\ee
where $e_a^{(0)}(x,u)$ is the solution of the EoM (a background), it is straightforward to show
\be \nonumber
\mathcal{F}_2^a(x,u) = \frac{\lambda_2}{2} \left( \{ g^{(0)}(x,u) \stackrel{\star}{,} h^a(x,u) \}
 +  \big\{ \{ e_b^{(0)}(x,u) \stackrel{\star}{,} h^b(x,u) \} \stackrel{\star}{,} e_{(0)}^a(x,u) \big\} + \mathcal{O}(h^2) \right) .
\ee
In case of the simplest type of background belonging to the geometric phase,
\be \nonumber
e_a^{(0)}(x,u) = e_a^{(0)\mu}(x)\, u_\mu
\ee
the contribution to the linearized EoM is at most second-order in spacetime derivatives. If the background is not of this type, it must have an infinite Taylor expansion in $u$ and as a consequence its contribution to the linearized EoM have an infinite number of terms with no bounds on order in spacetime derivatives.

Similarly, we can construct potential Lagrangian terms by taking higher polynomials in the MHS vielbein, all of them having the higher dimension in the geometric phase in $d>4$ than the terms already discussed. There are two interesting terms of a topological origin. If the number of spacetime dimensions is even, $d=2r$, there exists a Lorentz-scalar MHS tensor
\be \nonumber
P_r(x,u) = \epsilon^{a_1 b_1 \ldots a_r b_r}\, T_{a_1 b_1}(x,u) \star \cdots \star T_{a_r b_r}(x,u)
\ee
where $\epsilon^{a_1\ldots a_d}$ is the Levi-Civita symbol. As it is a generalization of the Chern term we refer to it as the MHS Chern tensor. It can be used to construct Lagrangian terms, the simplest being
\be
L_P(x,u) = \lambda_P\, P_r(x,u) \;.
\ee
Note that this term is parity-odd. In $d=4$ it has the same dimension as the MHSYM term. It is not hard to show that it is a topological term,
\be \label{mhsct}
P_r(x,u) = D^\star_{a_1} \left( \epsilon^{a_1 b_1 \ldots a_r b_r}\, e_{b_1}(x,u) \star T_{a_2 b_2}(x,u) \star \cdots \star T_{a_r b_r}(x,u) \right)
\ee
where we used (\ref{hsscp}) and (\ref{omricc2}). As a consequence it does not contribute to the (bulk) EoM, but may possibly lead to non-perturbative effects if the theory contains topologically non-trivial configurations analogous to instantons.

If the number of spacetime dimensions is odd, $d=2r+1$, we can construct the following Lagrangian term
\be
L_{\mathrm{CS}}(x,u) = \epsilon^{a b_1 c_1 \ldots b_r c_r}\,
 \{ e_a(x,u) \stackrel{\star}{,} T_{b_1 c_1}(x,u) \star \cdots \star T_{b_r c_r}(x,u) \} \;.
\ee
This tensor is parity-odd as well. From (\ref{mhsct}) it follows that it can be obtained as a boundary term from the MHS Chern term. It is thus natural to call it the MHS Chern-Simons tensor. It produces the following contribution to the EoM
\be
\mathcal{F}_\mathrm{CS}^a(x,u) = d\, \epsilon^{a b_1 c_1 \ldots b_r c_r}\,
 T_{b_1 c_1}(x,u) \star \cdots \star T_{b_n c_n}(x,u) 
\ee
which we call the MHS Cotton tensor. In $d=3$ the MHS Chern-Simons term has a lower dimension than the MHSYM term so it dominates in the IR regime. As the EoM of the pure MHS Chern-Simons theory in $d=3$ is
\be
\epsilon^{abc} T_{bc}(x,u) = 0 \qquad\Rightarrow\qquad T_{bc}(x,u) = 0
\ee
which shows that the MHSCS theory is topological.

\subsection{Conservation laws and conserved charges in MHSYM theory}
\label{ssec:conslc}

\subsubsection{Covariant vs.\ non-covariant conservation laws}
\label{sssec:cococl}

Here we would like to examine in more detail the question of conservation laws and conserved charges in MHS theories, taking MHSYM theory as an example. As is well-known in a Lorentz covariant theory a current satisfying the continuity equation on-shell (i.e., with EoM applied)\footnote{For equalities valid on-shell we use the symbol "$\doteq$". The equalities valid for generic field configurations satisfying proper boundary conditions are denoted by the simple equality sign "$=$".}
\be \label{scoeq}
\partial_\mu J^\mu(x) \doteq 0
\ee
encodes a local conservation of charge defined by
\be
Q_V(t) = \int_V d^{d-1}\mathbf{x}\, J^0(x) \;.
\ee
We refer to (\ref{scoeq}) as the conservation law. There are situations, especially when local symmetries are present, in which the total charge identically vanishes for all physical configurations. In the case of local symmetries such trivial charges are connected to the proper gauge transformations. Gauge transformations usually contain a subgroup of improper gauge transformations that are connected to non-trivial conserved charges. As the conserved charges in gauge theories can be written as asymptotic integrals, improper gauge symmetries are recognized by their "soft" fall-off in the limit $r\to\infty$. Here we are interested in extracting conservation laws and non-trivial charges in the framework of the MHS symmetry.

In theories with local symmetries covariant conservation laws appear naturally. In the case of MHS symmetry they are defined in the master space and are of the form
\be \label{mhscol}
\mathcal{D}_a^\star \mathcal{J}^a(x,u) \doteq 0
\ee
where $\mathcal{D}_a^\star$ is the MHS covariant derivative and $\mathcal{J}^a(x,u)$ is an MHS (tensor) current. An example is the matter current $J_\mathrm{m}(x,u)$, defined in (\ref{eomco}), which is the source in the MHS vielbein EoM. In ordinary theories with non-commutative local symmetries, such as YM theory and GR, a covariant conservation law does not directly imply conserved charges. Let us take as an example the matter energy-momentum tensor in GR, which is covariantly conserved. This does not imply conservation of the matter energy and momentum. We can construct the corresponding energy-momentum pseudotensor which is conserved in the sense of (\ref{scoeq}), but which also contains a contribution from the spin-2 sector. In the case of the MHS symmetry, the covariant conservation (\ref{mhscol}) automatically generates the conservation law (\ref{scoeq}). This is because the MHS covariant derivative is by the definition a Moyal commutator and every Moyal commutator is a total divergence in the master space
\be
0 \doteq \mathcal{D}_a^\star \mathcal{J}^a(x,u)
 = \partial^x_\mu A_{\mathcal{J}}^\mu(x,u) + \partial^\mu_u B^{\mathcal{J}}_\mu(x,u) \;.
\label{ccfccc}
\ee
Integrating both sides of (\ref{ccfccc}) over the auxiliary space and assuming that boundary terms in the auxiliary space are zero, we conclude
\be \label{phccur}
J^\mu(x) \equiv \int d^du \, A_{\mathcal{J}}^\mu(x,u)
\ee
is conserved. While covariantly conserved master field currents can usually be written in closed and compact expressions, we see from (\ref{ccfccc}) and the structure of the Moyal product that physically conserved spacetime currents $J^\mu(x)$ may have a rather involved and cumbersome form when written explicitly.

One example is the presence of the conserved matter charge in the MHS theory based on $U(1)$ spin-1 subgroup that is generated by the constant improper MHS transformation $\varepsilon(x,u) = \varepsilon = \mathrm{const}$. The MHS vielbein is neutral (invariant) under its action and so does not contribute to the $U(1)$ charge. It is the only conserved charge with such properties that is generated by the MHS symmetry. We now pass to a detailed study of conservation laws in the case of the MHSYM theory.

\subsubsection{Conserved currents from EoM}
\label{sssec:ccEoM}

In the geometric phase conserved charges are directly obtained from the EoM following the standard procedure used in ordinary YM theory and GR. Let us demonstrate this in the case of MHSYM theory coupled to the matter whose EoM is
\be
\frac{1}{g^2_\mathrm{ym}} \mathcal{D}^\star_b F^{ba}(x,u) \doteq \mathcal{J}_\mathrm{m}^a(x,u) \;.
\ee
We now use (\ref{gvieldef}) and move all nonlinear terms in $h_a(x,u)$ to the right hand side, obtaining
\be \label{ymme2}
\frac{1}{g^2_\mathrm{ym}} \partial^x_b F_{(1)}^{ba}(x,u) \doteq \tilde{\mathcal{J}}^a(x,u)
\ee
where
\be
F_{(1)}^{ab}(x,u) = \partial^x_a h_b(x,u) - \partial^x_b h_a(x,u)
\ee
and
\be
\tilde{\mathcal{J}}^a(x,u) &=& \mathcal{J}_\mathrm{m}^a(x,u) - \frac{i}{g^2_\mathrm{ym}} \left( 2 [h^b \stackrel{\star}{,} \partial^x_b h^a] - [h_b \stackrel{\star}{,} \partial_x^a h^b]
 + [\partial^x_b h^b \stackrel{\star}{,} h^a]  \right) -
\nonumber \\
&& -\, \frac{1}{g^2_\mathrm{ym}} \big[ h^b \stackrel{\star}{,} [h^a \stackrel{\star}{,} h_b] \big] \;.
\ee
Taking $a=0$ in (\ref{ymme2}) we get
\be \label{mhsgl}
\tilde{\mathcal{J}}^0(x,u) \doteq \frac{1}{g^2_\mathrm{ym}} \partial^x_j F_{(1)}^{j0}(x,u)
\ee
which is the MHS Gauss's law, while taking the spacetime divergence of Eq. (\ref{ymme2}) yields the continuity equation
\be \label{ceqtc}
\partial^x_a \tilde{\mathcal{J}}^a(x,u) \doteq 0
\ee
showing that the current $\tilde{\mathcal{J}}(x,u)$ is conserved in the master space (before integrating over auxiliary space). From Gauss's law (\ref{mhsgl}) it follows that the corresponding locally conserved charge can be written as a surface space integral
\be
\tilde{Q}_V(t,u) &=& \int_V d^{d-1}\mathbf{x}\, \tilde{\mathcal{J}}^0(x,u)
\noindent \\
&\doteq& \int_V d^{d-1}\mathbf{x}\, \partial^x_j F_{(1)}^{j0}(x,u)
\nonumber \\
&\doteq& \oint_{S(V)} d^{d-2}a_j\, F_{(1)}^{j0}(x,u)
\ee
which is Gauss's law in the integral form. Equation (\ref{ceqtc}) encodes a tower of conserved charges. To see this, we Taylor expand both sides in auxiliary coordinates around $u=0$ to obtain an infinite set of conserved charges
\be \label{cseom}
\tilde{Q}^{\mu_1\cdots\mu_n} \doteq \oint d^{d-2}a_j\, F_{(1)}^{j0\mu_1\cdots\mu_n}(x)
\ee
where
\be
F_{(1)}^{j0}(x,u) = \sum_{n=0}^\infty F_{(1)}^{j0\mu_1\cdots\mu_n}(x)\, u_{\mu_1} \cdots u_{\mu_n} \;.
\ee

\subsubsection{Conservation laws from the Noether method}
\label{sssec:emt}

Let us now construct conservation laws by applying the Noether method. For simplicity, we restrict ourselves to the pure MHSYM theory. First, using the MHSYM EoM 
\be \label{HSYMeom}
\qquad \mathcal{D}_a^\star T^{ab}(x,u) \doteq 0
\ee
we conclude that a generic on-shell variation of the MHSYM master Lagrangian can be written as
\be \label{Lymosv}
\delta L_\mathrm{ym}(x,u) \doteq -\frac{1}{2 g_{\mathrm{ym}}^2}
 \mathcal{D}^\star_a \{ T^{ab}(x,u) \stackrel{\star}{,} \delta e_b(x,u) \} \;.
\ee
If we take the variation to be an MHS variation
\be
\delta_\varepsilon e_a(x,u) = \mathcal{D}^\star_a \varepsilon(x,u)
\ee
we see that the Lagrangian transforms as an MHS tensor,
\be \label{Lymgv}
\delta_\varepsilon L(x,u) = i [ L(x,u) \stackrel{\star}{,} \varepsilon(x,u) ] \;.
\ee
We now use (\ref{Lymosv}) and (\ref{Lymgv}) to write
\be \label{ceqgen}
0 \doteq \frac{1}{2 g_{\mathrm{ym}}^2} \left( 
 \mathcal{D}^\star_a \big\{ T^{ab}(x,u) \stackrel{\star}{,} \mathcal{D}^\star_b \varepsilon(x,u) \big\}
 - \frac{i}{2} \big[ T_{ab}(x,u) \star T^{ab}(x,u) \stackrel{\star}{,} \varepsilon(x,u) \big] \right).
\ee
As both terms on the right hand side are Moyal commutators the equation has the form
\be
0 \doteq \partial^x_\mu A_\varepsilon^\mu(x,u) + \partial^\mu_u B^\varepsilon_\mu(x,u) \;.
\ee
Again, integrating over the auxiliary space and assuming that all boundary terms vanish, we obtain a standard conservation law (in the form of the continuity equation)
\be
\partial^x_\mu J_\varepsilon^\mu(x) \doteq 0 \qquad,\qquad
 J_\varepsilon^\mu(x) \equiv \int d^du\, A_\varepsilon^\mu(x,u) \;.
\ee
The corresponding conserved charges
\be
Q_\varepsilon = \int d^{d-1}\mathbf{x}\, J_\varepsilon^0(x)
\ee
are non-trivial only for a small class of MHS parameters corresponding to improper gauge transformations. It is expected that rigid variations $\varepsilon(x,u) = \varepsilon(u)$, which we can expand as
\be \label{mhsrv}
\varepsilon(u) = \sum_{n=0}^\infty \xi^{\mu_1\cdots\mu_n} u_{\mu_1} \ldots u_{\mu_n}
\ee
with $\xi^{\mu_1\cdots\mu_n}$ constant and completely symmetric, fall into this class. Let us now focus on them.\footnote{We do not claim that (\ref{mhsrv}) is a complete set of improper MHS transformations. Quite the contrary, the experience from Maxwell's theory and GR teaches us that there should be a much larger set of conserved charges. We leave the more complete analysis of asymptotic symmetries to future work.}  We have already analyzed the $n=0$ case, which does not affect MHS vielbein and so (\ref{ceqgen}) becomes trivial ($0\doteq 0$). Let us now consider $n\ge1$ cases, by first constructing covariantly conserved currents. For this we first have to find the covariantized form of the rigid MHS variation (\ref{mhsrv}), and the simplest way to do this is by replacing $u_a \to e_a(x,u)$,\footnote{We are motivated by Jackiw's covariantisation trick \cite{Jackiw1978}.}
\be \label{mhsrvc}
\varepsilon(x,u) = \sum_{n=0}^\infty \xi^{a_1\cdots a_n} e_{a_1}(x,u) \star \ldots \star e_{a_n}(x,u)
\ee
where $\xi^{a_1\cdots a_n}$ is a constant tensor with symmetries guaranteeing reality of the MHS parameter $\varepsilon(x,u)$.\footnote{For $\varepsilon(x,u)$ in (\ref{mhsrvc}) to be real it has to be expressible purely in terms of Moyal commutators and/or anticommutators. If the number of anticommutators is odd, the parameter $\xi^{a_1\cdots a_n}$ is imaginary. When $\xi^{a_1\cdots a_n}$ is completely symmetric the expression (\ref{mhsrvc}) is a covariantization of (\ref{mhsrv}).} Now we use this in (\ref{ceqgen}) where we want to write the second term on the right hand side as a covariant divergence (the first term is already in this form). We do this by using the identity
\be
\co{ A_1 \star \ldots \star A_n }{X} = \sum_{j=1}^n 
\co{ A_j }{A_{j+1} \star \ldots \star A_{n} \star X \star A_{1} \star \ldots \star A_{j-1}}
\ee
valid for generic master space functions $A_j(x,u)$ and $X(x,u)$, to write
\be
&& i \co{ e_{a_1} \star \ldots \star e_{a_n} }{ T_{ab} \star T^{ab} }
\nonumber \\
&&\qquad\qquad = \sum_{j=1}^n 
\mathcal{D}^\star_{a_j} \left({e_{a_{j+1}} \star \ldots \star e_{a_{n}} \star T_{ab} \star T^{ab} 
\star e_{a_{1}} \star \ldots \star e_{a_{j-1}}}\right). \qquad
\ee
Using this in (\ref{ceqgen}) we obtain
\be
\mathcal{D}^\star_a \mathscr{T}^a_\xi(x,u) \doteq 0
\ee
with the covariantly conserved currents given by
\be
\mathscr{T}^a_\xi &=& \xi^{b_1\cdots b_n} \bigg( 
 \ac{ T^{ac} }{
 \mathcal{D}^\star_c \left( e_{b_{1}} \star \ldots \star e_{b_{n}} \right)} +
 \0 \\ 
&& + \,\frac{1}{2}  \sum_{j=1}^n 
 \delta{}^a_{b_j}  {e_{b_{j+1}} \star \ldots \star e_{b_{n}} \star T_{cd} \star T^{cd} \star e_{b_{1}} \star \ldots \star e_{b_{j-1}}} \bigg).
\ee

The currents related to totally symmetric $\xi^{b_1\cdots b_n}$
\be
\mathscr{T}^a{}_{b_1\cdots b_n} &=& \ac{ T^{ac} }{
 \mathcal{D}^\star_c \left( e_{(b_{1}} \star \ldots \star e_{b_{n})} \right)} + 
 \0 \\ 
&& +\,\frac{1}{2}  \sum_{j=1}^n 
 \delta{}^a_{(b_j}  {e_{b_{j+1}} \star \ldots \star e_{b_{n}} \star T_{|cd|} \star T^{cd} \star e_{b_{1}} \star \ldots \star e_{b_{j-1})}} \qquad
\label{hsccc}
\ee
play a special role as they are obtained by covariantizing the rigid MHS symmetries. Another reason for its special status is that they have the softest behavior at spatial infinity ($r\to\infty$) in the geometric phase, which means that they are main candidates for producing non-trivial charges. Since the corresponding conserved charges are described by totally symmetric tensors, they should be related to the charges (\ref{cseom}) obtained by the previous method.

The $n=1$ case in (\ref{mhsrv}) corresponds to spacetime translations, leading to energy-momentum conservation,  and is therefore of special importance. Fixing $n=1$ in (\ref{hsccc}) we get the covariant master energy-momentum tensor 
\be \label{HSemd}
\mathscr{T}^a_b(x,u) = \{ T^{ac} \stackrel{\star}{,} T_{bc} \}
 - \frac{1}{2}\, \eta^a_b\, T_{cd} \star T^{cd}
\ee
which is symmetric, and in $d=4$ traceless. As expected, the obtained expression has the same form as in non-commutative field theories \cite{Abouzeid2001, DasFrenkel2003,Grimstrup2004,Balasin2015}.

\subsection{MHSYM as a matrix theory}
\label{ssec:ymmat}

There is another way to represent MHS theories discussed above, which uses the connection between the Moyal product and the Weyl-ordered operator product well known from the phase space formulation of a quantized particle. If we define the Hilbert space $\mathcal{H}$ with the complete set of operators $\hat{x}^\mu, \hat{u}_\mu$ satisfying commutation relations
\be
[ \hat{x}^\mu , \hat{u}_\nu ] = i \delta^\mu_\nu \quad,\quad [ \hat{x}^\mu , \hat{x}^\nu ] = 0 = [ \hat{u}_\mu , \hat{u}_\nu ]
\ee
there is a bijective map (for a fixed ordering scheme) between the set of linear operators on $\mathcal{H}$ and the set of functions on the master space, i.e.,\footnote{One usually refers to $O(x,u)$ as the symbol of operator $\hat{O}$.}
\be \label{weylmoyal}
\textrm{End}(\mathcal{H}) \ni \hat{O} \quad\longleftrightarrow\quad
 O(x,u) \in C^\infty (\mathcal{M} \times \mathcal{U} ) \;.
\ee
If one defines a product of two operators with the (symmetric) Weyl ordering of $x$ and $u$, its pull-back to the master space (through the map (\ref{weylmoyal})) defines the Moyal product of corresponding master space functions (symbols)
\be
\hat{O}_1\, \hat{O}_2 \quad\longleftrightarrow\quad
O_1(x , u) \star O_2(x , u) \;.
\ee
This map is such that the trace of an operator is given by the integral of the corresponding function over the master space
\be
\tr (\hat{O}) = \int d^dx\, \frac{d^du}{(2\pi)^d}\, O(x,u) \;.
\ee
Using this map it is now evident that all models for MHS theories can be written in this operator language, and therefore as a type of matrix models.\footnote{For an explicit proof that the Moyal product is a matrix product see \cite{Merkulov}.} For example, the MHSYM theory can be written as
\be
S_\mathrm{ym} = - \frac{(2\pi)^d}{4 g_{\mathrm{ym}}}\, \tr \left( [ \hat{e}_a\, , \hat{e}_b ] [ \hat{e}^a , \hat{e}^b ] \right)
\ee
where  $\hat{e}^a$ are operators on $\mathcal{H}$, components of a vector in the $SO(1,d-1)$ representation. The MHS symmetry is now represented by unitary linear operators
\be
\hat{U}_\mathcal{E} = \exp(-i\hat{\mathcal{E}})
\ee
which act on the MHS vielbein operator as
\be \label{mvtran}
\hat{e}_a \to \hat{U}_\mathcal{E}\, \hat{e}_a\, \hat{U}_\mathcal{E}^\dagger
\ee
with all operator products defined with symmetric Weyl ordering.

\section{Model building: matter sector}
\label{sec:mattHS}

Now we turn to the question of consistent coupling of matter fields to the MHS vielbein. We investigate three possible routes for achieving this.

\subsection{Minimal matter: spacetime fields}
\label{ssec:framelth}

There is a way to couple matter fields to the MHS vielbein that naturally accommodates the operator (matrix model) formulation introduced in section \ref{ssec:ymmat} and has the additional bonus of describing matter by purely spacetime fields. The latter means that we do not have to increase the number of degrees of freedom in the matter sector. This approach was originally introduced in \cite{Bekaert:2009ud,Bekaert:2010ky} for the case of the Klein-Gordon field and in \cite{BCDGPS,paper2} for the case of the Dirac field. Here we review these constructions in a generalised and compact manner (see also appendix \ref{app:Bekaert}).

The starting point in the construction is the observation, already noted in section \ref{ssec:ymmat}, that the MHS vielbein can be represented by a linear operator acting on a particular Hilbert space on which "position operators" $\hat{x}^\mu$ and their conjugate momenta $\hat{u}_\mu$ are represented. The matter configuration $\phi$ is represented by a state vector $| \alpha_\phi \rangle$ in this Hilbert space. Matter fields are then simply wave functions in the $x$-representation
\be
\phi_r(x) = \langle x , r | \alpha_\phi \rangle 
\ee
where the index $r$ allows for a non-trivial finite dimensional representation of the Lorentz group and possibly also of some internal group of symmetries. The classical action for a free matter field can simply be written as a particular expectation value
\be \label{fmfao}
S_m^{(0)}[\phi] = \langle \alpha_\phi | \hat{K}(\hat{u}) | \alpha_\phi \rangle
\ee
where $\hat{K}(\hat{u})$ is a linear operator depending on spins and masses of matter fields. For bosonic (fermionic) fields it is quadratic (linear) in $u$. In particular, for the complex Klein-Gordon field defined on flat spacetime,
\be \label{lopKG}
\hat{K}_\mathrm{s}(\hat{u}) = \eta^{ab} \hat{u}_a \hat{u}_b - m^2 
\ee
while for the Dirac field,
\be \label{lopD}
\hat{K}_D(\hat{u}) = - \gamma^0 (\gamma^a \hat{u}_a + m)
\ee
where $\gamma^a$ are Dirac matrices. 

As described in section \ref{ssec:ymmat}, the MHS transformations are represented by unitary linear operators 
$\hat{U}_\mathcal{E} = \exp(-i\hat{\mathcal{E}}(\hat{x},\hat{u}))$,
\be \label{mmftr}
| \alpha_\phi \rangle_\mathcal{E} = \hat{U}_\mathcal{E} | \alpha_\phi \rangle \;.
\ee
where $\hat{\mathcal{E}}^\dagger = \hat{\mathcal{E}}$, and where a symmetric ordering of $\hat{x}^\mu$ and $\hat{u}_\mu$ is assumed. The free field actions are invariant under the rigid MHS transformations for which $\hat{\mathcal{E}} = \hat{\mathcal{E}}(\hat{u})$, that lead to the well-known rigid HS variations (\ref{rhss}) after using Taylor expansion (\ref{gHsse}). The minimal way to have matter actions invariant under the local MHS transformations is to make the substitution
\be \label{rtoghss}
\hat{u}_a \to \hat{e}_a \quad\Longrightarrow\quad \hat{K}(\hat{u}) \to \hat{K}(\hat{e})
\ee
in (\ref{fmfao}) to obtain
\be \label{mcmao}
S_m[\phi,e] = \langle \alpha_\phi | \hat{K}(\hat{e}) | \alpha_\phi \rangle
\ee
which is the action for matter minimally coupled to the MHS vielbein. The (local) MHS symmetry is a straightforward consequence of (\ref{mvtran}) and (\ref{mmftr}).\footnote{Again, putting aside subtleties that appear for improper gauge transformations.} 

As before, we can write matter actions in the Moyal product language. As reviewed in appendix \ref{app:Bekaert}, using the phase space formalism we can write minimally coupled matter actions in the form 
\be \label{mcmact}
S_m[\phi,e] = \int d^dx\, d^du\, \Tr \big( W_\phi(x,u) \star K(e(x,u)) \big)
\ee
where the trace is performed over Lorentz and internal indices carried by matter fields and the Wigner function can be written as
\be
(W_\phi(x,u))_{rs} = \phi_r(x) \star \delta^d(u) \star \phi_s(x)^* \;.
\ee
By construction both the Wigner function and $K(e(x,u))$ are MHS tensors. It then follows that Lagrangians for minimally coupled matter, defined by (\ref{mcmact}), are also MHS tensors. Note that coupling the matter this way explicitly breaks the translational symmetry in the auxiliary space. The matter action is also formally defined for non-geometric configurations.

To understand the nature of the minimal coupling in MHS theory in the geometric phase, we first use (\ref{gvieldef}) to separate free and interacting parts of the action by writing
\be
K(e(x,u)) = K(u) + K_\mathrm{int}(h(x,u);u)
\ee
where the explicit dependence of $K_\mathrm{int}$ on $u$ is present only for bosonic fields. Coupling to the MHS potential is linear for fermionic matter and quadratic for bosonic matter. Substituting this into (\ref{mcmact}) we get
\be
S_m[\phi,h] = S_m^{(0)}[\phi] + S_m^{(\mathrm{int})}[\phi,h] 
\ee
where by definition $S_m^{(0)}$ is the action for the free field and the interaction term can be written as
\be
S_m^{(\mathrm{int})}[\phi,h] = \int d^dx\, d^du\, \big(\phi^*_r(x) \star K^{rs}_\mathrm{int}(x,u) \star \phi_s(x) \big) \delta^d(u) \;.
\label{Smintg}
\ee 

Let us demonstrate the above construction on two important examples of matter, Dirac and Klein-Gordon fields. In case of the complex Klein-Gordon field by using (\ref{lopKG}), (\ref{rtoghss}) and (\ref{gmetric}) we obtain
\be
K_\mathrm{int}(x,u) = h(x,u)
\ee
where $h(x,u)$ is a composite object obtained from the MHS potential by (\ref{hharel}). It was shown in 
\cite{Bekaert:2009ud,Bekaert:2010ky} that after using Taylor expansion (\ref{hsvte}) the interacting part of the action is given by
\be
S_m^{(\mathrm{int})}[\varphi,h] = \sum_{s=0}^\infty \int d^dx\,
 J^{(s)}_{\mu_1\cdots\mu_s}(x)\, h_{(s)}^{\mu_1\cdots\mu_s}(x) \;.
\ee
where the spin-$s$ currents are of the form
\be
J^{(s)}_{\mu_1\cdots\mu_s}(x) = \frac{i^s}{2^s}\, \varphi(x)^\ast \!\stackrel{\leftrightarrow}{\partial}_{\mu_1} \cdots
 \stackrel{\leftrightarrow}{\partial}_{\mu_s}\!\! \varphi(x) \;.
\ee
In case of the Dirac field $\psi(x)$ we have
\be \label{KintD}
K_\mathrm{int}(x,u) = - \gamma^0 \gamma^a h_a(x,u) \;.
\ee
Taylor expanding the MHS potential $h_a(x,u)$ as in (\ref{gvieldef}) one gets
\be \label{mDfc}
S_m^{(\mathrm{int})}[\psi,h] = \sum_{n=0}^\infty \int d^dx\,
 J^{a}_{(n)\mu_1\cdots\mu_n}(x)\, h_{a}^{(n)\mu_1\cdots\mu_n}(x)
\ee
where the HS currents (\cite{BCDGPS,paper2} are given by
\be
J^{a}_{(n)\mu_1\cdots\mu_n}(x) = \frac{i^n}{2^n}\, \bar{\psi}(x) \gamma^a \!\stackrel{\leftrightarrow}{\partial}_{\mu_1} \cdots \stackrel{\leftrightarrow}{\partial}_{\mu_n}\!\! \psi(x) \;.
\ee

Since we can write the Wigner function for off-diagonal matrix elements $\langle \alpha | (\cdots) | \beta \rangle$, this formalism can accommodate Lagrangian terms quadratic and non-diagonal in spacetime matter fields.

\subsection{Master field matter}
\label{ssec:mfmatt}

Another way to couple matter in an MHS symmetric way is to describe it by master fields, $\phi(x,u)$. This type of matter is necessary if one wants to introduce supersymmetry. The simplest representations are adjoint and fundamental.

\subsubsection{Adjoint representation}

Matter in the adjoint representation is described by MHS tensors, which means that the MHS covariant derivative is given by
\be
\mathcal{D}_a^\star \phi(x,u) = i [ e_a(x,u) \stackrel{\star}{,} \phi(x,u) ] \;.
\ee
In case of minimal coupling the action is then constructed in the standard way, by substituting 
$\partial_a^x \to \mathcal{D}_a^\star$. This type of matter shares some properties with the MHS gauge sector action: the master fields are real, actions are also defined in the non-geometric phases, and they can be written in the form of matrix models.

Let us apply this to the free Majorana spin-1/2 field $\psi(x,u)$. The MHS action for minimal coupling is\footnote{When discussing Majorana spinors, $d$ is assumed to be such that it allows for their existence.}
\be
S_M[\psi,e] = \frac{1}{2}\int d^dx\, d^du\, \bar{\psi}(x,u) \star (i\gamma^a \mathcal{D}_a^\star - M) \psi(x,u) \;.
\ee
In the operator formulation this is
\be
S_M[\psi,e] = - \frac{(2\pi)^d}{2}\, \tr \left( \bar{\hat{\psi}} \big( \gamma^a [ \hat{e}_a , \hat{\psi}] + M \hat{\psi} \big) \right).
\ee
Putting together the MHSYM action and the action for one minimally coupled massless Majorana field one gets the simplest supersymmetric MHS theory.

In case of the real scalar field the minimal coupling is described by the following action
\be
S_s[\varphi,h] = \int d^dx\, d^du\, \big[ \eta^{ab} (D^\star_a \varphi)^\ast \star D^\star_b \varphi
 - m^2 \varphi^\ast \star \varphi - V_\star(\varphi^\ast \star \varphi) \big]\:.
\ee

\subsubsection{Fundamental representation}

Matter in the fundamental representation of MHS symmetry transforms as
\be \label{hsvbm}
\phi_\mathcal{E}(x,u) = e_\star^{-i\mathcal{E}(x,u)} \star \phi(x,u)
\ee
from which it follows
\be
\phi_\mathcal{E}(x,u)^\ast = \phi(x,u)^\ast \star e_\star^{i \mathcal{E}(x,u)}
\ee
In the YM-like formalism the MHS covariant derivative in the fundamental representation is
\be \label{hscdf}
D_a^\star \phi = \partial^x_a \phi + i\, h_a \star \phi \;.
\ee
It is simple to check the MHS covariance
\be
(D_a^\star \phi)^\mathcal{E} = e_\star^{-i\mathcal{E}} \star D_a^\star \phi \;.
\ee

MHS invariants are constructed by Moyal-sandwiching MHS tensors between $\phi^*$ or $(D_a^\star \phi)^\ast$ from the left and $\phi$ or $D_a^\star \phi$ from the right. Using these invariants we can produce candidates for Lagrangian terms, with minimal coupling defined in the usual manner by a substitution of MHS covariant derivative for partial spacetime derivative in free field actions. 

However, minimal prescription based on (\ref{hscdf}) can be defined only in the geometric phase. In addition, it is not natural from the the perspective of a matrix model formulation. These shortfalls can be avoided by using the prescription 
\be \label{hscdm}
\partial_a^x \phi(x,u) \to i\, e_a(x,u) \star \phi(x,u) \;.
\ee
From the relation
\be
i\, e_a(x,u) \star \phi(x,u) = i u_a \phi(x,u) + \frac{1}{2} \partial_a^x \phi(x,u) + i\, h(x,u) \star \phi(x,u)
\ee
it is obvious that it differs from (\ref{hscdf}). To understand the origin of this degeneracy of minimal prescriptions, let us consider an example of the master Dirac field $\psi(x,u)$. In this case it is easy to show that
\be \label{Dlag}
- \bar{\psi} \gamma^a \star e_a \star \psi = \frac{i}{2} \bar{\psi} \gamma^a \star D^\star_a \psi
 - \frac{i}{2} \overline{D^\star_a \psi} \star \gamma^a \psi + u_a \bar{\psi} \gamma^a \star \psi \;.
\ee
The first two terms on the right hand side produce the master Lagrangian kinetic term which one would obtain by the minimal coupling prescription based on (\ref{hscdf}), leading to the action 
\be \label{Dactc}
S_{D1}[\psi,e] = \int d^dx\, d^du\, \bar{\psi}(x,u) \star \big(i \gamma^a D^\star_a -  M \big) \psi(x,u) \;.
\ee
On the left hand side of (\ref{Dlag}) is the expression which takes natural matrix model form when used in the action 
\be \label{Dactm}
S_{D2}[\hat{\psi},\hat{e}] = - \Tr \left( \hat{\bar{\psi}} (\gamma^a \hat{e}_a + M) \hat{\psi} \right)
\ee
and is formally defined for all phases of the MHS theory (it also takes care of hermicity by automatism). The two actions differ already at the free field level, i.e., for $h_a(x,u)=0$. We now see that the difference between Lagrangians in two prescriptions is the second term on the right hand side of (\ref{Dlag}) which is an MHS scalar. Its existence is a consequence of the fact that Lagrangian terms for matter in the fundamental representation are MHS scalars, which means that they can be multiplied by functions of the auxiliary coordinates without breaking any of the important symmetries.\footnote{Of course we should be careful not to break symmetries which we would like to preserve, such as Lorentz symmetry and translations in spacetime. The symmetry under translations in the auxiliary space is broken by such multiplications, but it is not obvious why we should try to protect this symmetry.}

Let us mention that mater fields in the fundamental representation have an additional peculiarity that the rigid MHS variations with $n=1$ ($s=2$) act differently than in the case of the MHS vielbein and previously discussed realizations of matter,
\be
\delta_{\varepsilon_{(1)}} \phi &=& -i\, \varepsilon^\mu u_\mu \star \phi
\nonumber \\
&=&  -i\, \varepsilon^\mu u_\mu\, \phi - \frac{\varepsilon^\mu}{2}\, \partial^x_\mu \phi \;.
\ee
We see that it does not describe spacetime translations. One consequence is that the MHS transformations in this case can be consistently truncated only to the lowest spin sector ($n=0$) when master fields are Taylor-expanded around $u=0$.

\section{Low-spin sector}
\label{sec:lows}

\subsection{Emergent geometry in the MHS theory}
\label{ssec:interpr12}

We have seen in section \ref{ssec:framelth} that the minimal way to incorporate interacting matter inside the MHS framework leads to the picture in which matter perceives spacetime fields obtained by Taylor expanding the MHS vielbein in the auxiliary space
\be \label{hsvtef}
e_a(x,u) = \sum_{n=0}^\infty e_a^{(n)\mu_1\ldots\mu_n}(x)\, u_{\mu_1} \cdots u_{\mu_n}
\ee
as a HS background. From this viewpoint the lowest two components, $e_a^{(1)}(x)$ and $e_a^{(1)\mu}(x)$, play the roles of the $U(1)$ potential and the emergent spacetime vielbein, respectively. In this section we focus on the low-spin sector $n\le1$ ($s\le2$), with the goal of finding out what sort of spacetime geometry emerges from the MHS framework. 

For this reason we write the expansion (\ref{hsvtef}) in the form
\be \label{s2trunc}
e_a(x,u) = A_a(x) + E_a{}^\mu(x)\, u_\mu + \ldots
\ee
and similarly for the MHS variation parameter
\be \label{e2trunc}
\varepsilon(x,u) = \epsilon(x) + \varepsilon^\mu(x)\, u_\mu + \ldots
\ee
and in all expressions ignore higher spin components (with $n>1$) denoted above by ellipses. We noted before that the truncation to the low-spin sector is apparently consistent at the level of MHS symmetry and EoM. However, we should keep in mind that such truncated configurations are not physical (see Sec\ \ref{ssec:physcon}), so our findings based on this truncation are of limited importance.

If $B(x,u)$ and $C(x,u)$ are generic master fields, their Moyal bracket truncated to low spin sector,
\be
i [ B(x,u) \stackrel{\star}{,} C(x,u) ] &=&
 \frac{\partial C(x,u)}{\partial x^\mu} \, \frac{\partial B(x,u)}{\partial u_\mu}
 - \frac{\partial B(x,u)}{\partial x^\mu} \, \frac{\partial C(x,u)}{\partial u_\mu} + \ldots
\nonumber \\
 &=& - \{ B(x,u) , C(x,u) \}_\mathrm{PB} + \ldots
\label{mbs12}
\ee
is given by the Poisson bracket, where the master space plays the role of the phase space. From (\ref{mbs12}) it follows that the set of spin-2 truncated master fields is closed under the Moyal bracket. The Taylor expansion of (\ref{mbs12}) is given by
\be
i [ B(x,u) \stackrel{\star}{,} C(x,u) ] &=& B_{(1)}^\nu(x)\, \partial_\nu C_{(0)}(x)
 - C_{(1)}^\nu(x)\, \partial_\nu B_{(0)}(x)
\nonumber \\
 && + \left( B_{(1)}^\nu(x)\, \partial_\nu C_{(1)}^\mu(x)
 - C_{(1)}^\nu(x)\, \partial_\nu B_{(1)}^\mu(x) \right) u_\mu + \ldots
\\
&=& \pounds_{B_{(1)}} C_{(0)} - \pounds_{C_{(1)}} B_{(0)} + \big(\pounds_{B_{(1)}} C_{(1)} \big)^\mu\, u_\mu
+ \ldots
\label{mbt12l}
\ee
The last line is obtained by recognizing the differential-geometric structure, with Lie derivatives treating $B_{(0)}(x)$ and $C_{(0)}(x)$ as scalar fields and $A_{(1)}^\mu(x)$ and $B_{(1)}^\mu(x)$ as vector fields on the spacetime manifold. We will see below that this is generally true in our construction -- all expressions truncated to the low-spin sector ($s\le2$) are going to be diff-covariant.

Let us apply this to the MHS variation of the MHS vielbein (\ref{esvha}). Using (\ref{s2trunc}), (\ref{e2trunc}) and (\ref{mbt12l}) we obtain that the low-spin spacetime fields transform as 
\be
\delta_\epsilon A_a(x) &=& \pounds_{E_a} \epsilon(x)
\label{hsvgs1} \\
\delta_\varepsilon A_a(x) &=& - \pounds_{\varepsilon} A_a(x)
\label{hsvgs2} \\
\delta_\epsilon E_a{}^\mu(x) &=& 0
\label{hsvvs1} \\
\delta_\varepsilon E_a{}^\mu(x) &=& \big(\pounds_{E_a} \varepsilon \big)^\mu(x)
\label{hsvvs2}
\ee
We now see that the MHS variation with $n=1$ acts as an infinitesimal diffeomorphism defined by
\be \label{idifs2}
x'^\mu = x^\mu + \varepsilon^\mu(x)
\ee
under which $E_a{}^\mu(x)$ behaves as a set of vector fields, while $A_a(x)$ behaves as a set of scalars. Assuming that the frame $E_a{}^\mu(x)$ is regular, i.e., there exists a co-frame $E^a{}_\mu(x)$ satisfying
\be
E^a{}_\mu(x) E_a{}^\nu(x) = \delta_\mu^\nu \qquad,\qquad E^b{}_\mu(x) E_a{}^\mu(x) = \delta_a^b
\ee
an MHS variation with $n=0$ acts on $A_\mu(x) = E^a{}_\mu(x)\, A_a(x)$ as
\be
\delta_\epsilon A_\mu(x) = - \partial_\mu \epsilon(x)
\ee
while the frame $E_a{}^\mu(x)$ is invariant. Taken all together, $E_a{}^\mu(x)$ can be identified as the (inverse) vielbein, while $A_a(x)$ can be identified as a $U(1)$ gauge potential vector field in the non-coordinate basis of the vielbein. The $n=0$ MHS variations are infinitesimal $U(1)$ gauge transformations, while $n=1$ MHS variations are infinitesimal diffeomorphisms. Note that this interpretation is only valid in the geometric phase, in which the frame $E_a{}^\mu(x)$ invertible, to which we now turn our attention. 

A word of caution is necessary here. If we keep higher spin contributions, the diff-covariant structure, at least as defined in the standard way, is apparently lost. This can be traced to the mixing (or twisting) of the HS transformations noted already at the level of the rigid transformations in section \ref{ssec:mhsstr}. The effects of twisting can be seen by analysing $n=1$ finite (large) MHS transformations of MHS tensors, 
\be
\mathcal{E}(x,u) = \mathcal{E}^\mu(x)\, u_\mu
\ee
where for the sake of simplicity we assume that the $n=0$ component of the MHS tensor is vanishing 
\be
X_{a\ldots}(x,u) = X_{a\ldots}^{(1)\mu}(x)\, u_\mu + \mathcal{O}(u^2)
\ee
From the definition (\ref{hslarge}) and the Baker-Campbell-Hausdorff formula it follows that the spacetime vector field $X_{a\ldots}^{(1)\mu}(x)$ transforms as
\be \label{lHSdiff}
\big(X_{a\ldots}^{(1)\mathcal{E}}\big)^\mu = \big(\exp(\pounds_{\mathcal{E}}) X_{a\ldots}^{(1)}\big)^\mu + \ldots \;.
\ee
This should be compared with the diff-transformation of a vector field
\be
V'^\mu(x') = \partial_\nu \zeta^\mu(x)\, V^\nu(x) \qquad,\qquad x'^\mu = \zeta^\mu(x)
\ee
The large MHS transformation (\ref{lHSdiff}) is a diffeomorphism, where the connection between parameter fields seems to be given by
\be
\zeta^\mu(x) - x^\mu &=& \sum_{r=0}^\infty \frac{(\mathcal{E}(x) \cdot \partial)^r}{(r+1)!}\, \mathcal{E}^\mu(x)
\nonumber \\
&=& \frac{e^{\mathcal{E}(x) \cdot \partial} - 1}{\mathcal{E}(x) \cdot \partial}\, \mathcal{E}^\mu(x) \;.
\label{diffmhsc}
\ee 
We have checked this relation up to quartic order \cite{xAct,xTras}. We see that large MHS transformations in spin-2 sector are indeed finite diffeomorphisms, but that the naturally defined parameters of the two descriptions are related in a complicated way given by (\ref{diffmhsc}). 

Let us now analyze the metric, whose low-spin components in the MHS framework are naturally obtained from (\ref{gvmconn}). The result is
\be
&& g_{(0)}(x) = \frac{1}{2} A_a(x)\, A^a(x) + \frac{1}{2} \partial_\nu E_a{}^\mu(x)\, \partial_\mu E^{a\nu}(x) + \ldots
\label{gs120} \\
&& g_{(1)}^\mu(x) = E^{a\mu}(x)\, A_a(x) + \ldots
\label{gs121} \\
&& g_{(2)}^{\mu\nu}(x) = E^{a\mu}(x)\, E_a{}^\nu(x) + \ldots \;.
\label{gs122}
\ee
Relations (\ref{gs121})-(\ref{gs122}) confirm the identification of 
\be
A^\mu(x) \equiv E^{a\mu}(x)\, A_a(x)
\ee
as a $U(1)$ vector potential, $E_a{}^\mu(x)$ as a vielbein and 
\be
g^{\mu\nu}(x) \equiv E^{a\mu}(x)\, E_a{}^\nu(x)
\ee
as the (inverse) metric tensor. The terms on the right hand side of (\ref{gs120}) are responsible for producing the seagull interaction terms when a Klein-Gordon matter field is minimally coupled to the MHS field.

To understand the induced geometry, we have to find the induced linear connection. The natural way to obtain it is by analysing the $n=1$ component of the MHS covariant derivative of an MHS tensor. Using (\ref{hsscp}) and (\ref{mbt12l}) we get
\be \label{hscd2}
(\mathcal{D}^\star_a B_{b\ldots})_{(1)}^\mu(x) = \big(\pounds_{E_a} B_{b\ldots}^{(1)} \big)^\mu(x) + \ldots \qquad
\ee
from which it follows that the induced covariant derivative of a vector field should be given by
\be \label{hsconn2}
(\nabla_{E_a} V)^\mu \equiv E_a{}^\nu \nabla_\nu V^\mu = \big(\pounds_{E_a} V \big)^\mu \;.
\ee
Multiplying by $E^a{}_\mu(x)$ we finally obtain
\be
\nabla_\nu V^\mu &=& E^a{}_\nu \big(\pounds_{E_a} V \big)^\mu
\nonumber \\
&=&  \partial_\nu V^\mu + E_a{}^\mu \partial_\rho E^a{}_\nu\, V^\rho \;.
\ee
This means that the MHS symmetry induces the following linear connection
\be \label{hslconn}
\Gamma^\mu{}_{\rho\nu} = E_a{}^\mu \partial_\rho E^a{}_\nu = - E^a{}_\nu \partial_\rho E_a{}^\mu \;.
\ee
The obtained linear connection is very much different from the Levi-Civita connection. For one, the torsion tensor is generally non-vanishing, as it can explicitly be checked
\be
T^\mu{}_{\rho\nu} &=& \Gamma^\mu{}_{\nu\rho} - \Gamma^\mu{}_{\rho\nu}
\nonumber \\
&=& \xi^\mu{}_{\rho\nu} \equiv E^a{}_\rho E^b{}_\nu \xi^\mu{}_{ab}
\label{hsitt}
\ee
where $\xi^\mu{}_{ab}(x)$ is
\be \label{cunhd}
\xi^\mu{}_{ab} = \big(\pounds_{E_a} E_b \big)^\mu = \xi^c{}_{ab} E_c{}^\mu \;.
\ee
$\xi^c{}_{ab}(x)$ are known as coefficients of anholonomy. As a consistency check, let us calculate the $n=1$ component of the HS torsion. It is easy to show that it is given by
\be \label{hsc2}
T_{ab}^{(1)\mu}(x) = \xi^\mu{}_{ab}(x) + \ldots
\ee
which is consistent with (\ref{hsitt}).

Also, the linear connection (\ref{hslconn}) is not metric compatible, the nonmetricity tensor being
\be
Q_\rho{}^{\mu\nu} \equiv \nabla_{\!\rho}\, g^{\mu\nu}
 = T^{\mu}{}_{\rho\sigma}\, g^{\rho\nu} + T^{\nu}{}_{\rho\sigma}\, g^{\rho\mu}
 = T^{\mu\nu}{}_{\sigma} + T^{\nu\mu}{}_{\sigma}
\ee
which is generally non-vanishing. Note that the nonmetricity tensor is not independent but is fully (algebraically) expressible in terms of torsion. The same is true for the Riemann tensor for which it can be shown that it can be expressed in terms of the torsion tensor and its covariant derivatives.

Let us also calculate the spin connection in the geometry induced by the MHS construction. The simplest way to find it is to use
\be
A^a{}_{b\mu} = E^a{}_\nu\, \nabla_\mu E_b{}^\nu \;.
\ee
Using (\ref{hslconn}) we get
\be
A^a{}_{b\mu} = E^a{}_\nu E_b{}^\rho\, T^\nu{}_{\mu\rho} = T^a{}_{\mu b} \;.
\ee
We see that $A_{ab\mu}$ is not antisymmetric in its first two indices, which is a manifestation of metric incompatibility.  Again, we see that the induced spin connection is fully determined by the torsion.

\subsection{Connection to teleparallelism}
\label{ssec:telgrav}

We have seen that the induced spacetime geometry found in the $s=2$ ($n=1$) sector of MHS theory seems rather unusual. The linear connection is metric-incompatible, and both the torsion and the Riemann tensor are non-vanishing. In fact, it is closely related to teleparallel geometry. The key observation is that there is only one independent fundamental tensor, the torsion, and all others are expressible in terms of it.

Let us first briefly review the concept of distant parallelism or teleparallelism.\footnote{For a detailed exposition of teleparallel geometry and gravity see the book \cite{AldPer}.} Let us assume that a differentiable manifold is equipped with a linear connection $\Gamma_+$, which is not symmetric. Teleparallelism is a requirement on the linear connection that there exists a frame of vector fields (an inertial frame) $E_a{}^\mu(x)$ that globally satisfies 
\be \label{tpcond}
\nabla^+_\mu E_a{}^\sigma \equiv \partial_\mu E_a{}^\sigma + \Gamma^\sigma_{+\rho\mu} E_a{}^\rho = 0 \;.
\ee
From the definition of the covariant derivative it follows that the linear connection is given by
\be \label{Wbconn}
\Gamma^\sigma_{+\rho\mu} = E_a{}^\sigma \partial_\mu E^a{}_\rho = - E^a{}_\rho\, \partial_\mu E_a{}^\sigma
\ee
which is known as the Weitzenb\"{o}ck connection. If the metric is naturally defined by taking the inertial frame as the vielbein
\be
g^{\mu\nu} = \eta^{ab} E_a{}^\mu E_b{}^\nu
\ee
then from (\ref{tpcond}) it obviously follows that the Weitzenb\"{o}ck connection is metric compatible
\be
\nabla^+_\rho g_{\mu\nu} = 0
\ee

An outstanding property of the Weitzenb\"{o}ck connection is that its corresponding spin (Lorentz) connection is vanishing
\be \label{tscv}
\mathscr{A}^a_{+b\mu} = E^b{}_\sigma \nabla_\mu E_a{}^\sigma = 0
\ee
for inertial frames. Inertial frames are related to one another through global Lorentz transformations,
\be
E_a{}^\mu(x) \to \Lambda_a{}^b E_b{}^\mu(x) \;.
\ee
Note that by performing a {\it local} Lorentz transformation
\be
E_a{}^\mu(x) \to \Lambda_a{}^b(x) E_b{}^\mu(x) \qquad,\qquad \partial_\mu \Lambda_a{}^b \ne 0
\ee
one passes to a non-inertial frame which does not satisfy (\ref{tpcond}). As a consequence the spin connection in the transformed frame is non-vanishing but still trivial (i.e. flat), 
\be
\mathscr{A}^a_{+b\mu} \to \Lambda_b{}^c \partial_\mu \Lambda_c{}^a \;.
\ee

It follows directly from (\ref{tscv}) that the Riemann tensor also vanishes
\be
R^a_{+b\mu\nu} = 0
\ee
which, as a consistency check, one could also show using the Weitzenb\"{o}ck linear connection. This means that, beside the metric, the only nontrivial fundamental tensor in teleparallel geometry is torsion, which is given (using the inertial frame) by
\be \label{tgtor}
T^\mu_{+\nu\rho} &\equiv& \Gamma^\mu_{+\rho\nu} - \Gamma^\mu_{+\nu\rho}
\nonumber \\ 
&=& - \xi^\mu{}_{\nu\rho}
\ee
where the anholonomy $\xi$ was defined in (\ref{cunhd}).

The simplest Lagrangians of teleparallel gravity theories are of the form \cite{NewGR}
\be \label{atpg}
S_{\mathrm{tg}} = \int d^dx\, E \left( c_1\, T^\rho_{+\mu\nu} T_\rho^{+\mu\nu}
 + c_2\, T^\rho_{+\mu\nu} T_+^{\nu\mu}{}_\rho + c_3\, T^\rho_{+\mu\rho} T^+_{\nu\mu}{}^\nu  \right)
\ee
It can be shown that if one takes 
\be \label{cttg}
c_1 = \frac{1}{4} \qquad,\qquad c_2 = \frac{1}{2} \qquad,\qquad c_3 = - 1
\ee
then (\ref{atpg}) becomes equal, up to a boundary term, to the Einstein-Hilbert action. The theory based on this action is usually called the teleparallel equivalent of General Relativity (TEGR). One of the advantages of the teleparallel formulation (over the Einstein-Hilbert one) is having a manifestly diff-covariant Lagrangian that is first order in derivatives. Teleparallel gravity theories were first studied by Albert Einstein already in 1920's \cite{AE1928}.

Let us now finally connect the geometry emerging from the MHS symmetry with teleparallel geometry. Comparing their respective linear connections, (\ref{hslconn}) and (\ref{Wbconn}), we see that
\be
\Gamma^\mu{}_{\nu\rho} = \Gamma^\mu_{+\rho\nu}
\ee
i.e., our linear connection is the {\it opposite} of the teleparallel one. It is then not strange that torsions are related by
\be \label{hsttp}
T^\mu{}_{\nu\rho} = - T^\mu_{+\nu\rho} \;.
\ee
This means that the covariant derivative induced by MHS symmetry can be written in terms of the covariant derivative of  teleparallel geometry
\be \label{hscdtp}
\nabla = \nabla_+ - T_+ \;.
\ee
Using (\ref{hsttp}) and (\ref{hscdtp}) we can express any covariant expression in the teleparallel geometry as a covariant expression in the opposite of teleparallel geometry. As a special case, it means that a manifestly covariant EoM in the emergent MHS geometry can be expressed as a manifestly covariant EoM in teleparallel geometry, and vice versa. This will be important below in the discussion of $s=2$ sector of EoM in the MHSYM model.

Teleparallel gravity can be obtained by gauging the group isometric to the group of spacetime translations \cite{Hayashi:1967se,Pereira:2019woq,LeDelliou:2019esi}. As the global MHS transformations have such a subgroup ($n=1$ sector), it is not surprising that there is a connection between the MHS theory and teleparallel gravity.

\subsection{MHSYM model in the $s\le2$ sector}
\label{ssec:YMs12}

In view of the preceding discussion on the induced spacetime geometry in the MHS construction, it is interesting to study the $s\le2$ sector of EoM of the MHSYM model. We put (\ref{s2trunc}) into the MHSYM EoM (\ref{HSYMeom}) and take into consideration only the purely $s\le 2$ components. The $s=1$ component of the EoM is given by
\be \label{ym121c}
0 = E_b{}^\nu \partial_\nu F^{ba} - \xi^{\nu ba} \partial_\nu A_b + \ldots
\ee
where $\mathbf{F}$ is the 2-form field strength of the spin-1 $U(1)$ spacetime vector potential 1-form $\mathbf{A}$, i.e.
\be
F_{ab} &=& E_a{}^\mu E_b{}^\nu F_{\mu\nu} 
= E_a{}^\mu E_b{}^\nu (\partial_\mu A_\nu - \partial_\nu A_\mu)
\nonumber \\
&=& E_a{}^\nu \partial_\nu A_b - E_b{}^\nu \partial_\nu A_a + A_c\, \xi^c{}_{ab} \;.
\label{F2def}
\ee
The $s=2$ component of the EoM is
\be \label{ym122c2}
0 = E_b{}^\nu \partial_\nu \xi^{\mu ba} - \xi^{\nu ba} \partial_\nu E_b{}^\mu + \ldots \;.
\ee
After some manipulation we can rewrite this equation in the equivalent form
\be \label{ym122c}
0 = E_c{}^\nu \partial_\nu \xi^{abc} - \xi^a{}_{cd}\, \xi^{cdb} + \ldots \;.
\ee
Also, using (\ref{ym122c}) and (\ref{F2def}) we can write the $s=1$ EoM component (\ref{ym121c}) as
\be \label{ym121c2}
0 = E_b{}^\nu \partial_\nu F^{ba} - \xi^{bca} F_{bc} + \ldots
\ee
which is manifestly $U(1)$-gauge invariant.

Terms denoted by ellipses, and also complete $s>2$ components of EoM, vanish if all $s>2$ components of the MHS vielbein vanish. This is a consequence of the fact that the MHSYM theory can be consistently truncated to the low-spin ($s\le2$) sector at the level of EoM.\footnote{We remind the reader that such truncated configurations are not physically acceptable in our formalism.}

Let us rewrite the low-spin EoM (\ref{ym122c})-(\ref{ym121c2}) within the framework of teleparallel geometry by using (\ref{hsttp}) and the fact that in teleparallel gravity Lorentz connection is vanishing in inertial frames so the Lorentz covariant derivative is simply the coordinate derivative
\be
\mathscr{D}^+_\mu = \partial_\mu \qquad\Longrightarrow\qquad \mathscr{D}^+_a = E_a{}^\mu \partial_\mu \;.
\ee
Using all of this we can write (\ref{ym121c2}) as
\be \label{ymets1}
\mathscr{D}_b^+ F^{ba} + T_+^{bca} F_{bc} = 0
\ee
which is now fully diff- and  $U(1)$ covariant. Similarly, (\ref{ym122c}) becomes
\be \label{ymets2}
\mathscr{D}^+_c T_+^{abc} + T^a_{+cd}\, T_+^{cdb} = 0
\ee
which is manifestly diff-covariant. We stress that in this form all objects in (\ref{ymets1}) and (\ref{ymets2}) should be calculated using the Weitzenb\"{o}ck connection. In other words, they are formally written within the realm of teleparallel gravity, and not the geometry induced by the MHS symmetry. 

There is an interesting piece of history here. The equation (\ref{ymets2}) was first written by Albert Einstein in 1929, with a motivation to unify electromagnetism with gravity \cite{AE1929}.\footnote{For this reason he added by hand one more equation to EoM in an attempt to project out unwanted degrees of freedom. Already in 1930 he abandoned this attempt.}. Einstein observed that it is not possible to write the diff-covariant action which produces (\ref{ymets2}) as its EoM inside the realm of the teleparallel geometry, because the left hand side is not covariantly conserved (in particular it does not belong to the class of theories defined in (\ref{atpg})). It is amusing that we obtained EoM (\ref{ymets2}) from an action principle by truncating MHSYM theory.

\section{Conclusions}
\label{app:concl}

We have investigated a novel way of gauging higher derivative (or higher spin) rigid symmetries present in all free field theories in flat spacetime, originally proposed in \cite{paper1,paper2}. The proposal is based on the symmetries observed for relativistic matter fields coupled linearly to an infinite tower of higher-spin fields \cite{Bekaert:2009ud,Bekaert:2010ky,BCDGPS}. The important feature of this proposal, which we refer to as Moyal-higher-spin (MHS), is that the gauge potential is intrinsically (off-shell) defined on the master space, which is a direct product of the ordinary spacetime and an auxiliary space of the same dimensionality. The Moyal product introduces a particular type of non-commutativity which acts between the spacetime and the auxiliary space, whereas spacetime commutativity is kept. We have shown that one can construct actions for the MHS gauge potential and matter coupled to it by mimicking the standard Yang-Mills procedure, providing in this way such desired properties as Poincare symmetry, a perturbatively stable vacuum, the formal application of BRST quantization rules, mass generation by spontaneous symmetry breaking, supersymmetrization and $L_\infty$ structure. 
We have shown that the gauge potential can be manifestly covariantized, in a way which has resemblances to the teleparallel gravity construction based on gauging the group of translations. A consequence of this generalisation, which we call geometry-like formulation is an appearance of new phases, including the strongly coupled with an unbroken vacuum. It also provides a way of casting the MHS theory into an (infinite dimensional) matrix model form.

An obvious question is how to connect the MHS construction to the standard Wigner classification of irreducible representations of the Poincare group. On the level of equations of motion, the MHS Yang-Mills theory can be written in terms of an infinite tower of tensor spacetime fields with unbounded rank, thus making contact with higher spin field theories. The spin-2 sector has a differential geometric interpretation, closely related to teleparallel geometry, with a linear connection opposite to Weizenb\"{o}ck's. The MHS theory formally allows for more phases, even with a singular emergent tangent frame (vielbein). The YM phase corresponds to a regular emergent geometry , while the unbroken phase presents the maximally singular (i.e. vanishing) frame. However, when comparing with the standard higher spin program there are important differences. Higher spin fields obtained in the spacetime MHS decomposition are not independent in the usual sense, one consequence being that there is no regular off-shell purely spacetime and manifestly Lorentz covariant description of the MHS theory. This points at the possibility that the MHSYM theory is not compatible with the Fock space composed of weakly interacting particles with finite spins. One can now understand how the intrinsic master space nature of the MHS construction surpasses the barriers raised by no-go theorems.

Interestingly, a manifestly covariant off-shell description of MHS theory in terms of an infinite tower of higher-spin spacetime fields is possible in the Euclidean regime. This description can be used to analyse amplitudes by using Feynman diagrams in the standard fashion. We postpone the perturbative analysis of amplitudes to our forthcoming paper and report some preliminary results. Tree level amplitudes with all external lines belonging to the minimal spacetime matter described in section \ref{ssec:framelth} are zero for nontrivial momenta (they are proportional to delta functions). In the case of master field matter, the tree-level amplitudes are softened by form factors with the UV fall-off faster than any power, when compared to QED.  

There are some obvious questions arising from our analysis. The fact that the MHS theory is based on gauging rigid higher spin symmetries and allows for higher-spin descriptions under some circumstances raises questions as to its possible connection to the tensionless limit of string field theory. That it has a natural matrix model formulation suggests possible connections with ideas on higher-spin constructions proposed by Steinacker \cite{Steinacker:2019fcb,Steinacker:2020xph}. Another question is whether the MHS theory, in the present or modified form, has any connections to the infinite (or continuous) spin representations of the Poincare group, which also contain a continuous (but compact) "index" \cite{Bekaert:2017khg,Schuster:2013pta}. We leave these and other questions to our future work.

\acknowledgments


\noindent
P.D.P.\ would like to thank Erwin Schroedinger Institute for Mathematics and Physics (University of Vienna) for support during the visit under the framework of ESI Programme and Workshop "Higher Spins and Holography", and Evgeny Skvortsov and Per Sundell for stimulating discussions. S.G is also grateful to Dario Francia, Per Sundell, Massimo Taronna, Konstantin Alkalaev, Alexey Sharapov, Evgeny Skvortsov, Xavier Bekaert, Euihun Joung and to the organizers and participants of the workshop “Higher Spin Gravity: Chaotic, Conformal and Algebraic Aspects” at the Asia Pacific Center for Theoretical Physics for discussions and comments on topics relevant to the paper. We especially thank Loriano Bonora as this work evolved from our mutual collaboration and for useful discussions and comments on early versions of the manuscript.This research has been supported by the University of Rijeka under the project uniri-prirod-18-256. The research of S.G.\ has been supported by the Israel Science Foundation (ISF), grant No.\ 244/17.

\appendix

\section{Properties of the Moyal product}
\label{app:moyal}

We use the following definition for the Moyal product of functions on the master
space $\mathcal{M} \times \mathcal{U}$ with coordinates $(x,u)$:
\be
a(x,u) \star b(x,u) &=& a(x,u)\,
\exp\!\left[\frac{i}{2}\left(\stackrel{\leftarrow}{\partial}_x \cdot
\stackrel{\rightarrow}{\partial}_u
 - \stackrel{\rightarrow}{\partial}_x \cdot \stackrel{\leftarrow}{\partial}_u
\right)\right] b(x,u)
\nonumber \\
&=& \exp\!\left[\frac{i}{2}\left(\partial_x \cdot \partial_w - \partial_y \cdot \partial_u \right)\right] a(x,u)\, b(y,w)
 \Big|_{y=x\,,\, w=u} \;.
\ee
The partial derivatives are defined in the usual way, i.e.,
\be
\partial^x_\mu = \frac{\partial}{\partial x^\mu} \quad, \quad \partial_u^\mu = \frac{\partial}{\partial u_\mu} \;. 
\ee
The Moyal product is associative
\be \label{mpass}
\big( a(x,u) \star b(x,u) \big) \star c(x,u) = a(x,u) \star \big( b(x,u) \star
c(x,u) \big)
\ee
and Hermitian under the complex conjugation
\be \label{ccmoyp}
(a(x,u) \star b(x,u))^\ast = b(x,u)^\ast \star a(x,u)^\ast \;.
\ee

The Moyal product can, for a class of functions with well behaved fall of conditions, be calculated in the integral form
\begin{equation}
\label{Moyal_integral}
a(x,u) \star b(x,u) = \int d^dy\, d^d z\, \frac{d^dv}{(2\pi)^d}\frac{d^dw}{(2\pi)^d}
e^{i(yw - zu)}a(x+\frac{y}{2}, u + v) b(x+\frac{z}{2}, u + w)
\end{equation}
and a very useful and convenient way to make calculations with the Moyal product is by promoting coordinates to operators, of which we note one possible way:
\begin{equation}
\label{Moyal_operator}
a(x,u) \star b(x,u) = a(x,{\bf u})b(x,{\bf u}^\dag)
\end{equation}
with 
\begin{equation}
{\bf u} = u -\frac{i}{2}\vec{\partial}_x, \qquad  {\bf u}^\dag = u +\frac{i}{2}\overleftarrow{\partial}_x
\end{equation}

Note that the Moyal commutator of real functions is purely imaginary, while the Moyal anticommutator of real functions is real. It is important to note that the Moyal commutator obeys the Jacobi identity 
\be \label{mbji}
[ a \stackrel{\star}{,} [ b \stackrel{\star}{,} c ]] + [ c \stackrel{\star}{,} [ a
\stackrel{\star}{,} b ]]
 + [ b \stackrel{\star}{,} [ c \stackrel{\star}{,} a ]] = 0
\ee
and the derivation (or Leibniz) property
\be \label{derprop}
[a \stackrel{\star}{,} b \star c] = [a \stackrel{\star}{,} b] \star c + b \star [a \stackrel{\star}{,} c] \;.
\ee
The same properties are obeyed by the ordinary (matrix) commutator. From
(\ref{derprop}) it follows
\be
\{ [ a \stackrel{\star}{,} b] \stackrel{\star}{,} a \} = [ a \star a \stackrel{\star}{,} b ] \;.
\ee
The Moyal product satisfies the adjoint property under integration
\be
\int d^dx\, d^du\, \big( a(x,u) \star b(x,u) \big) c(x,u) &=& \int d^dx\, d^du\,
a(x,u) \big( b(x,u) \star c(x,u) \big)
\nonumber \\
 &=& \int d^dx\, d^du\, b(x,u) \big( c(x,u) \star a(x,u) \big)
\ee
where $a$, $b$ and $c$ are square-integrable functions on the master space. If we put
$c(x,u)=1$ we obtain
\be \label{mpintp}
\int d^dx\, d^du\, a(x,u) \star b(x,u) = \int d^dx\, d^du\, a(x,u)\, b(x,u) + \mbox{(boundary terms)} \;.
\ee

It is convenient to define deformations on standard functions on the master space by using Moyal instead of ordinary product in the Taylor expansion. We denote such functions with the $\star$ subscript, e.g., 
\be \label{estar}
e_\star^{a(x,u)} = \sum_{n=0}^\infty \frac{1}{n!}\, a(x,u)^{\star n}
\ee
where $a(x,u)^{\star n}$ is the Moyal product with $n$ factors of $a(x,u)$
\be
a(x,u)^{\star n} = a(x,u) \star a(x,p) \star \ldots \star a(x,u) \;.
\ee

\section{Proof of the theorem on triviality}
 \label{app:omtrivproof}
 
Here we prove that the HS field strength measures the triviality of HS configurations, i.e.
\be \label{hsctc}
\mbox{HS master space field is pure gauge} \qquad \Longleftrightarrow \qquad F_{ab}(x,u) = 0
\ee
in a domain of configurations containing $h_a(x,u)=0$.

\smallskip
\noindent
\emph{Proof.} 
From (\ref{omhps}) it follows directly that the theorem is valid in the linear approximation, since the linear term in (\ref{omhst}) can be interpreted as the exterior derivative of a form $h_a^{\mu_1\ldots\mu_n}(x)$ if Greek indices $\mu_j$ are treated as internal. It means that in the linearized theory if and only if $F_{ab}(x,u)=0$ in a ball, there exists a master space function $\varepsilon(x,u)$ such that
\be \label{lsolom0}
h_a(x,u) = \partial_a\, \varepsilon(x,u)
\ee
in the same ball. But this is just the linearised pure gauge condition for the MHS potential $h_a(x,u)$.

Let us extend this to large fields. First we prove the left-to-right arrow in (\ref{hsctc}). If the MHS vielbein field is pure gauge, then by (\ref{hslarge}) we can write it as
\be
e_a(x,u) = e_\star^{-i\, \mathcal{E}(x,u)} \star u_a \star e_\star^{i\, \mathcal{E}(x,u)}
 = u_a - i e_\star^{-i\, \mathcal{E}(x,u)} \star  \partial_a^x e_\star^{i\, \mathcal{E}(x,u)}
\ee
so a pure gauge MHS master field $h_a(x,u)$ is of the form
\be \label{hpuga}
h_a(x,u) =  -i e_\star^{-i\, \mathcal{E}(x,u)} \star  \partial_a^x e_\star^{i\, \mathcal{E}(x,u)} \;.
\ee
Plugging this into (\ref{omhps}), and using the identities
\be
e_\star^{-i\, \mathcal{E}(x,u)} \star e_\star^{i\, \mathcal{E}(x,u)} = 1 \qquad,\qquad
e_\star^{-i\, \mathcal{E}(x,u)} \star  \partial_a^x e_\star^{i\, \mathcal{E}(x,u)} = 
- \partial_a^x e_\star^{-i\, \mathcal{E}(x,u)} \star e_\star^{i\, \mathcal{E}(x,u)}
\ee
we conclude that MHS field strength $F_{ab}(x,u)$ vanishes for pure gauge HS fields. Therefore,
\be
\mbox{HS phase space field is pure gauge} \qquad \Longrightarrow \qquad F_{ab}(x,u) = 0 \;.
\ee

Proving the opposite direction of (\ref{hsctc}) happens to be more involved. We want to find the general solution of the equation
\be \label{ome0}
F_{ab}(x,u) = 0 \;.
\ee
To do this, let us start from the linearized solution (\ref{lsolom0}) and build a full solution by a formal perturbative series\footnote{In this construction it is not assumed that the HS potential $h_a(x,u)$ is small. We can introduce a formal parameter $\theta$, and consider (\ref{Om0solp}) as an expansion in $\theta$. Eventually we put $\theta \to 1$.}
\be \label{Om0solp}
h_a(x,u) = \sum_{n=1}^\infty \Delta^{(n)}_a(x,u) \;.
\ee
Introducing (\ref{Om0solp}) into (\ref{ome0}), using (\ref{omhps}), and collecting the terms of the same order, we obtain
\be \label{Om0ieq}
\partial^x_a \Delta^{(n)}_b(x,u) - \partial^x_b \Delta^{(n)}_a(x,u)
 = -i \sum_{r=1}^{n-1} [ \Delta^{(r)}_a(x,u) \stackrel{\star}{,} \Delta^{(n-r)}_b(x,u) ] \;.
\ee
We see that it has a form which can be attacked by mathematical induction. For $n=1$ it becomes
\be
\partial^x_a \Delta^{(1)}_b(x,u) - \partial^x_b \Delta^{(1)}_a(x,u) = 0
\ee
for which the general solution is
\be
\Delta^{(1)}_a(x,u) = \partial^x_a \mathcal{E}(x,u)
\ee
where $\mathcal{E}(x,p)$ is an arbitrary function. For $n=2$ we get
\be
\partial^x_a \Delta^{(2)}_b - \partial^x_b \Delta^{(2)}_a
 &=& -i\, [ \Delta^{(1)}_a \stackrel{\star}{,} \Delta^{(1)}_b ] 
\nonumber \\
 &=& -i\, [ \partial^x_a \mathcal{E}  \stackrel{\star}{,} \partial^x_b \mathcal{E} ]
\nonumber \\
 &=& -\frac{i}{2} \left( \partial^x_a [\mathcal{E}  \stackrel{\star}{,} \partial^x_b \mathcal{E} ] 
 - \partial^x_b [\mathcal{E}  \stackrel{\star}{,} \partial^x_a \mathcal{E} ] \right)
\ee
for which the solution is
\be
\Delta^{(2)}_b(x,u) = -\frac{i}{2}\, [\mathcal{E}(x,u)  \stackrel{\star}{,} \partial^x_a \mathcal{E}(x,u) ]
 + \partial^x_a \mathcal{E}'(x,u) \;.
\ee
The trivial exact part of the solution, which appears at every order, is of the same form as the first-order solution -- it introduces nothing new and can therefore be ignored in the construction of the general solution. Now we conjecture that the generic solution is given by
\be \label{Om0soln}
\Delta^{(n)}_a  &=& \frac{(-i)^{n-1}}{n!}\, [ \mathcal{E} \stackrel{\star}{,} [ \mathcal{E} \stackrel{\star}{,} \ldots
 [ \mathcal{E} \stackrel{\star}{,} \partial_a \mathcal{E} ] ] \ldots] \qquad (n-1\;\; \mbox{Moyal brackets})
\nonumber \\
 &=& \frac{(-i)^n}{n!}\, [ \mathcal{E} \stackrel{\star}{,} [ \mathcal{E} \stackrel{\star}{,} \ldots
 [ \mathcal{E} \stackrel{\star}{,} u_a ] ] \ldots] \qquad\quad (n\;\; \mbox{Moyal brackets}) \;.
\ee
This can be proved by induction. Using (\ref{Om0soln}) in (\ref{Om0solp}) gives us finally
\be
h_a(x,u)  &=& \sum_{n=1}^\infty \frac{(-i)^n}{n!}\, [ \mathcal{E}(x,u) \stackrel{\star}{,}
 [ \mathcal{E}(x,u) \stackrel{\star}{,} \ldots [ \mathcal{E}(x,u) \stackrel{\star}{,} u_a ] ] \ldots] 
\nonumber \\
 &=& e_\star^{-i\, \mathcal{E}(x,u)} \star u_a \star e_\star^{i\, \mathcal{E}(x,u)} - u_a
\nonumber \\
 &=& -i\, e_\star^{-i\, \mathcal{E}(x,u)} \star \partial^x_a e_\star^{i\, \mathcal{E}(x,u)} 
\ee
where during the passage from the first to the second line we used the Baker-Campbell-Hausdorff lemma. We have obtained (\ref{hpuga}), therefore we proved that
\be
F_{ab}(x,u) = 0  \qquad \Longrightarrow \qquad \mbox{HS potential is a pure gauge}
\ee
in some neighbourhood of $h_a(x,u) = 0$.  $\blacksquare$

\smallskip

The perturbative construction outlined above has some limitations, in the sense that one cannot obtain all configurations with vanishing field strength by gauge transforming the configuration $h_a(x,u) = 0$, which is 
\be \label{flatM}
e^{(0)}_a(x,u) = u_a = \delta_a^\mu u_\mu\;. 
\ee
To see this let us consider configurations of the form
\be \label{ctevac}
e_a(x,u) = M_a{}^\mu u_\mu
\ee
with $M$ a constant real $d\times d$ matrix, for which (\ref{ome0}) is also true. If we want to connect such configurations with the vacuum (\ref{flatM}) by an MHS transformation, the MHS gauge parameter must be of the form
\be
\mathcal{E}_\Lambda(x,u) = x^\mu \Lambda_\mu{}^\nu u_\nu
\ee
with $\Lambda$ again a constant real $d\times d$ matrix. Now one has
\be
[ \mathcal{E}_\Lambda(x,u) \stackrel{\star}{,} u_\mu ] = i \Lambda_\mu{}^\nu u_\nu
\ee
from which follows
\be
e^\Lambda_a(x,u) &\equiv& e_\star^{-i \mathcal{E}_\Lambda(x,u)} \star u_a \star e_\star^{i \mathcal{E}_\Lambda(x,u)}
\nonumber \\
&=& \sum_{n=0}^\infty \frac{(-i)^n}{n!}\, [ \mathcal{E}_\Lambda(x,u) \stackrel{\star}{,}
 [ \mathcal{E}_\Lambda(x,u) \stackrel{\star}{,} \ldots [ \mathcal{E}_\Lambda(x,u) \stackrel{\star}{,} u_a ] ] \ldots]
\nonumber \\
 &=& \delta_a^\mu\, (e^{\Lambda})_\mu{}^\nu u_\nu
\label{gecvac}
\ee
If $M$ cannot be written as an exponential of some matrix, the corresponding configuration (\ref{ctevac}) is not MHS gauge equivalent to the vacuum (\ref{flatM}).

There is more than one gauge orbit consisting of solutions of EoM satisfying $T_{ab}(x,u) = 0$, which is to say there are many vacua in the (pure) MHSYM theory. The above construction of the orbit of $h_a(x,u)=0$ was limited by taking $e^{(0)}_a(x,u) = \delta_a^\mu u_\mu$ as the $0^\mathrm{th}$-order term. If we had started with 
$e^{(0)}_a(x,u) = M_a{}^\mu u_\mu$, with matrix $M$ of rank $< d$,we would instead cover a different gauge orbit.

\section{Weyl-Wigner formalism}
\label{app:Bekaert}

Borrowing the language from the phase space formulation of a quantized particle, we can rewrite free field actions using the Wigner function, and show how a gauging of the MHS symmetry leads to the existence and transformation properties of the MHS potential. In a similar way, this was originally realized in \cite{Bekaert:2009ud, Bekaert:2010ky}, and in  \cite{BCDGPS} for the frame formalism.

\subsection{Wigner function}

One can rewrite the free field action for a (for simplicity massless) complex scalar field in the following way (as displayed in sect. (\ref{ssec:framelth}) and using $\hat{K}_\mathrm{s}(\hat{u}) = \eta^{ab} \hat{u}_a \hat{u}_b $)
\begin{equation}
S^{(0)}[\phi] = \int d^dx\, \partial_\mu \phi\, \partial^\mu \phi^\dag
 = \langle \alpha_\phi | \hat{K}_s(\hat{u}) | \alpha_\phi \rangle = \mathrm{Tr}\left(\hat{K}_s(\hat{u}) | \alpha_\phi \rangle\langle \alpha_\phi |\right) .
\end{equation}
Under the Weyl-Wigner map, a trace of a product of two operators is mapped to an integral over the phase space of a Moyal product of their respective symbols. The necessary symbols are calculated using the Wigner map
\be
\mathcal{W}[\hat{F}]=f(x,u) = \int d^d q\,\langle x - \frac{q}{2}| \hat{F}| x + \frac{q}{2}\rangle e^{i  q \cdot u}
\ee
providing the symbol of the kernel
\be
 \mathcal{W}[\hat{K}]=\int d^d q\,\langle x - \frac{q}{2}| \eta^{ab} \hat{u}_a \hat{u}_b| x
  + \frac{q}{2}\rangle e^{i  q \cdot u} = u^2
\ee
and the symbol of the projector (which turns out to be the Wigner function \cite{Wigner1932,Fairlie})
\be
\mathcal{W}\left[| \alpha_\phi \rangle\langle \alpha_\phi |\right] &=& \int d^d q\,\langle x - \frac{q}{2}| \alpha_\phi \rangle\langle \alpha_\phi | x + \frac{q}{2}\rangle\, e^{i  q \cdot u}
\nonumber \\ 
 &=&  \int d^d q\, \phi(x-q/2) \phi^\dagger(x+q/2)\, e^{i  q \cdot u}
\nonumber \\
&=& (2\pi)^d \phi(x) \star \delta^d(u) \star \phi^\dag(x)
\ee
where $\star$ is the Moyal product. The expression in the last line can be proved in the following way
\begin{align}
(2\pi)^d \phi(x) \star \delta^d(u) \star \phi^\dag(x) &= \int d^d q\,\phi(x) \star e^{iqu} \star \phi^\dag(x) 
\nonumber\\
& =\int d^d q\,\phi(x) \star\left[ e^{iqu} \phi^\dag(x +\frac{q}{2})\right] 
\nonumber \\
&=\int d^d q\,\phi(x - \frac{q}{2})\, e^{iqu} \phi^\dag(x +\frac{q}{2}) \;.
\end{align}
To avoid having factors of $(2\pi)$ in expressions, we redefine the Wigner function as 
\be
W_\phi \equiv (2\pi)^{-d} \mathcal{W}\left[| \alpha_\phi \rangle\langle \alpha_\phi |\right]
= \phi(x) \star \delta^d(u) \star \phi^\dag(x) \;.
\ee
By using (\ref{mmftr}), it can be shown that the Wigner function transforms as an MHS tensor in the adjoint representation
\begin{equation}
\label{Wigner_MHS}
\delta_\varepsilon W_\phi (x,u) = i[W_\phi(x,u)\stackrel{\star}{,}\varepsilon(x,u)] \;.
\end{equation}
Alternatively, the Wigner function can be thought of as the fundamental object used to describe matter, and its transformation properties can be postulated as above.

\subsection{Second order/metric formalism}

Using the Wigner function, the action for a free complex scalar field can be written out in the phase space as
\begin{equation}
S = \int d^dx\, \partial_\mu \phi\, \partial^\mu \phi^\dag = \int d^dx\, d^du\, u^2 \star W_\phi
\end{equation}
and we can see that for a rigid gauge parameter $\varepsilon(u)$ we have a symmetry of the action
\footnote{To see this, add and subtract $W_\phi(x,u) \star u^2\star \varepsilon(u)$ under the integral.}
\begin{align}
\delta S= i \int d^dx\, d^du\left([\varepsilon(u) \stackrel{\star}{,} W_\phi(x,u) \star u^2] +  W_\phi(x,u) \star [ u^2\stackrel{\star}{,}\varepsilon(u)]\right)=0 \;.
\end{align}
The first term is a boundary term and is discarded. The second term has a Moyal commutator of only $u$-dependent quantities, so it vanishes.

Considering a local symmetry $\varepsilon=\varepsilon(x,u)$, the variation of free field action is non-vanishing
\begin{equation}
\delta S =i\int d^dx\, d^du\, W_\phi(x,u) \star [ u^2\stackrel{\star}{,}\varepsilon(x,u)] \;.
\end{equation}
In the spirit of Yang Mills theory, this calls for a compensating field to obtain local invariance 
\begin{equation}
u^2 \to u^2 - h(x,u) \;.
\end{equation}
To first order in changes in $W_\phi$ and $h$, by neglecting total derivatives under the integral, we have
\begin{align}
\delta S =\int d^dx\, d^du\, W_\phi\star(- \delta h + i [(u^2 - h)\stackrel{\star}{,}\varepsilon]) \;.
\end{align}
To keep the action symmetric under local transformations, we can infer that $h(x,u)$ must transform as
\begin{align}
\delta h(x,u) &=2 u \cdot\partial^x \varepsilon(x,u) - i[h(x,u)\stackrel{\star}{,}\varepsilon(x,u)]
\end{align}
reproducing the result in \cite{Bekaert:2010ky}.

\subsection{First order/frame formalism}

The free scalar field action can be written in an equivalent way (due to the cyclicity of the Moyal product under integration) leading to the first order or frame representation.
\begin{equation}
S = \int d^dx\, \partial_\mu \phi\, \partial^\mu \phi^\dag = \int d^dx\, d^d u \, u^a \star W_\phi \star u_a \;.
\end{equation}
Again, for a rigid $\varepsilon(x,u) = \varepsilon(u)$ we have a symmetry since\footnote{To see this add and subtract $+ W_\phi\star \varepsilon \star u_a \star u^a - W_\phi\star \varepsilon\star u_a\star u^a + \varepsilon \star u_a \star W_\phi \star u^a - \varepsilon\star u_a \star W_\phi\star u^a$ and use cyclicity of the Moyal product under integration.}
\begin{align*}
\delta S &=i \int d^dx\, d^d u\, ( [ u^a \stackrel{\star}{,} W_\phi \star \varepsilon \star u_a]
 + [\varepsilon \stackrel{\star}{,} u^a]\star \{W_\phi \stackrel{\star}{,} u_a\}) \;.
\end{align*}
The first term is discarded as it is a boundary term, and the second term vanishes since $\varepsilon = \varepsilon(u)$.

In case where the gauge parameter $\varepsilon=\varepsilon (x,u)$ is also a function of $x$, the second term does not vanish, so again a compensating field is introduced, however this time it is a Lorentz vector field $h_a(x,u)$.
\begin{equation}
S = \int d^dx\, d^d u\, (u_a + h_a) \star W_\phi \star (u^a + h^a) \;.
\end{equation}
The variation of the action becomes:
\begin{align}
\delta S =& \int d^dx\, d^d u\, \Big( \delta h_a  \star \{ W_\phi \stackrel{\star}{,} (u^a + h^a)\} + i (u_a + h_a)\star W_\phi \star \varepsilon \star (u^a + h^a) \\&- i (u_a + h_a) \star \varepsilon\star W_\phi \star (u^a + h^a)\Big) \;.
\end{align}
Following the logic above and discarding the boundary terms we obtain:
\begin{align*}
\delta S = \int d^dx\, d^d u\, & (\delta h_a + i [\varepsilon\stackrel{\star}{,} (u_a + h_a)] ) \star
 \{ W_\phi \stackrel{\star}{,} (u^a + h^a)\}
\end{align*}
from which we conclude that the action is locally invariant if the compensating field has the following infinitesimal transformation law:
\begin{equation}
\delta h_a(x,u) = \partial_a \varepsilon(x,u) + i[h_a(x,u)\stackrel{\star}{,}\varepsilon(x,u)] \;.
\end{equation}
This reproduces the known result obtained using a free fermion action \cite{BCDGPS}.


\end{document}